\documentclass[11pt,a4paper]{article}

\usepackage{ascmac}
\usepackage{amsmath}
\usepackage{amssymb}
\usepackage{bm}
\usepackage[footnotesize,bf]{caption}
\usepackage{fullpage}
\usepackage{color}
\usepackage{float}
\usepackage{graphicx}
\usepackage[hidelinks]{hyperref} 
\usepackage[]{natbib}
\usepackage{subfigure}
\usepackage{txfonts}
\usepackage{threeparttable}
\usepackage{wrapfig}
\usepackage[]{multicol}
\usepackage{wrapfig}
\usepackage{authblk}

\usepackage{floatflt}

\title{\large Complete list of the ASTRO-H Science Working Group}
\date{\vspace{-0.5cm}}
\newcommand{\MakeWhitePaperTitle}{
	\begin{center}
		\begin{figure}
			\vspace{1cm}
			\begin{center}
				\includegraphics[width=0.2\hsize]{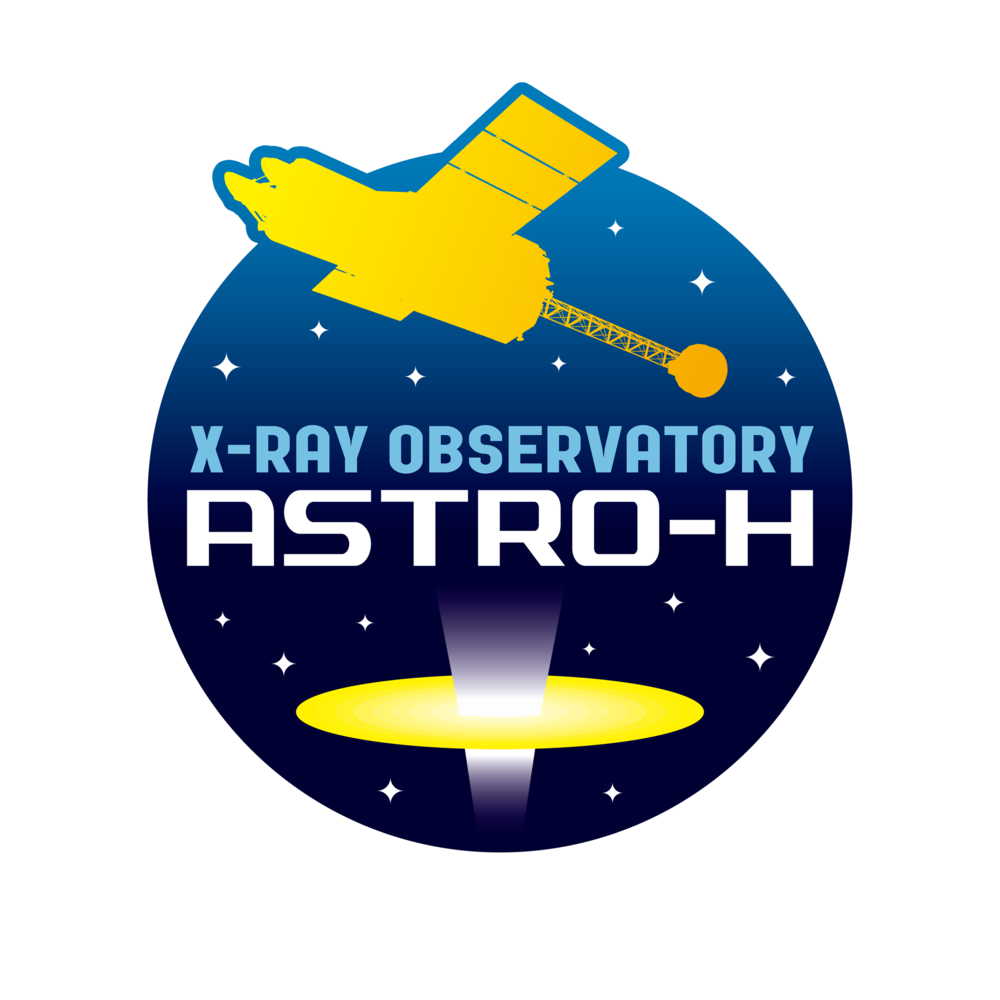}
			\end{center}
		\end{figure}
		\vspace{1cm}
		{\LARGE
		ASTRO-H Space X-ray Observatory\\
		White Paper\\
		}
		\vspace{5mm}
		{\large
		\WhitePaperTitle\\
		}
		\vspace{1cm}
		{
		\WhitePaperAuthors\\
		on behalf of the ASTRO-H Science Working Group
		}
	\end{center}
}

\usepackage{authblk}
\author[a]{Tadayuki~Takahashi}
\author[a]{Kazuhisa~Mitsuda}
\author[b]{Richard~Kelley}
\author[c]{Felix~Aharonian}
\author[d]{Hiroki~Akamatsu}
\author[e]{Fumie~Akimoto}
\author[f]{Steve~Allen}
\author[g]{Naohisa~Anabuki}
\author[b]{Lorella~Angelini}
\author[h]{Keith~Arnaud}
% ...
\affil[a]{Institute of Space and Astronautical Science (ISAS), Japan Aerospace Exploration Agency (JAXA), Kanagawa 252-5210, Japan}
\affil[b]{NASA/Goddard Space Flight Center, MD 20771, USA}
\affil[c]{Astronomy and Astrophysics Section, Dublin Institute for Advanced Studies, Dublin 2, Ireland}
\affil[d]{SRON Netherlands Institute for Space Research, Utrecht, The Netherlands}
\affil[e]{Department of Physics, Nagoya University, Aichi 338-8570, Japan}
\affil[f]{Kavli Institute for Particle Astrophysics and Cosmology, Stanford University, CA 94305, USA}
\affil[g]{Department of Earth and Space Science, Osaka University, Osaka 560-0043, Japan}
\affil[h]{Department of Astronomy, University of Maryland, MD 20742, USA}
% ...

\begin{document}

%---------------------------------------------
% title
%---------------------------------------------
\newcommand{\WhitePaperTitle}{Stars -- Accretion, Shocks, Charge Exchanges and Magnetic Phenomena}
\newcommand{\WhitePaperAuthors}{
	Y.~Tsuboi~(Chuo~University), K.~Ishibashi~(Nagoya~University), M.~Audard~(Universit\`{e} de Gen\'{e}ve), 
	K.~Hamaguchi~(UMBC/NASA), ~M.~A.~Leutenegger~(UMBC/NASA), 
	Y.~Maeda~(JAXA), K.~Mori~(Miyazaki~University), H, Murakami~(Rikkyo~University),
	Y.~Sugawara~(Chuo~University), and M.~Tsujimoto~(JAXA)
}
\MakeWhitePaperTitle

\input{macro.inp}
\newcommand{\etacar}{$\eta$~Car\ }
\newcommand{\ah}{{\it ASTRO-H}}

\begin{abstract}
X-ray emission from stars has origins as diverse as the stars
  themselves: accretion shocks, shocks generated in wind-wind
  collisions, or release of magnetic energy. Although the scenarios
  responsible for X-ray emission are thought to be known, the physical
  mechanisms operating are in many cases not yet fully understood.
  Full testing of many of these mechanisms requires high energy
  resolution, large effective area, and coverage of broad energy
  bands. The loss of the X-ray calorimeter spectrometer on board
  {\it ASTRO-E2} was a huge blow to the field; it would have provided a
  large sample of high resolution spectra of stars with high
  signal-to-noise ratio. Now, with the advent of the {\it ASTRO-H} Soft
  X-ray Spectrometer and Hard X-ray Imager, we will be able to examine
  some of the hot topics in stellar astrophysics and solve outstanding
  mysteries.
\end{abstract}

%---------------------------------------------
% member list
%---------------------------------------------
\maketitle
\clearpage

%---------------------------------------------
% TOC
%---------------------------------------------
\tableofcontents
\clearpage

%%%%%%%%%%%%%%%%%%%%%%%%%%%%%%%%%%%%%%%%%%%%%%%%%%%%%%%%%%%%
%\setcounter{section}{-1}
\section{X-ray Stellar Astrophysics} 
%\author{Prepared by K. Ishibashi}
%%%%%%%%%%%%%%%%%%%%%%%%%%%%%%%%%%%%%%%%%%%%%%%%%%%%%%%%%%%
%\setcounter{subsection}
%\subsection{Stars}

For understanding the universe and its evolution it is imperative to
understand the physics of stellar phenomena, as stars are one of the
most fundamental constituents of the universe. Stars have a major
influence on the dynamical state of surrounding media and enriches
them chemically with a fresh injection of heavy metals through their
slow or rapid evolution. Stars eventually evolve into other exotic
objects, such as white dwarfs, neutron stars, or a black hole. In the
process, stars may explode, as in supernova explosions. \hfill
\medskip

\noindent
But what happens if our understanding of stars is not adequate?
\hfill
\medskip

\noindent
The physical models and mechanisms describing exotic objects are based
on the best knowledge of stellar physics. If the foundation of our
understanding for stars stands on shaky ground, then astrophysics -- and
cosmology -- may be in peril.
\smallskip

And we understand them very little: how does a flare occur in stars?
Even magnetic activity on the Sun is not fully understood. How does
coronal heating occur in non-solar-type objects (or is it identical)?
What is the accretion geometry in very young stars? What density does
a pre-natal cocoon of a young star typically have?
\smallskip
% the physics of stars are multifaceted; it involves a variety                    
% of physical phenomena like accretion, shocks, release of                        
% magnetic free energy (i.e., reconnection), charge exchanges                     
% etc.                                                                            

In this section, we focuses on three key aspects of the \ah\
mission: high energy resolution ($5-7$~eV), highly accurate energy determination
(i.e., stable gain) of Soft X-ray Spectrometer (SXS) and its broad-band
capability combined with Soft and Hard X-ray Imagers (SXI and HXI).
\smallskip

For instance on flaring stars, the SXS provides sufficient energy
resolution to construct reliable differential emission measure at
the high-temperature end using Fe and Ni K emission lines; at the same time,
the HXI will allow the same flare to be probed at energies greater than
10~keV and finally to place a constraint on the low cut-off energy
in accelerated electrons, fully accounting the amount of energy
released by a flaring mechanism (hence leading to better understanding
of the mechanism itself).
\smallskip

Much like flaring stars, interactions of massive stars' winds can be
probed as well with the combination of SXI $+$ SXS $+$ HXI to reveal
the physical state of stellar wind shocks. Furthermore, for the first time,
\ah\ may be able to detect inverse Compton emission in its hard X-ray
tail and to reveal its connection with known thermal radiation seen
in them.
\smallskip

A traditional method of line diagnostics is valid as well for studies in
classical T-Tauri stars. At a higher energy (E $\ge$ 1 keV), the advantage
exists in the SXS to probe local densities ($10^{11}$ to $10^{13}~\mathrm{cm}^{-3}$)
associated with regions in which the emission of Ne IX and Mg XI triplets emerge.
This will enable us to probe X-ray emission of young stars for which
accretion plays a major role.

\smallskip
For studies of protostars and diffuse X-ray emission surrounding star
formation region, the two key aspects of the SXS become crucial:
high energy resolution and highly accurate energy determination 
for dynamic studies.
A rotating core of a protostar is so deeply embedded in its prenatal cocoon
and cannot be probed directly. But the shape and centroid position of He-like
and fluorescent Fe line tells the story about its origin in an accreting disk.
A dynamic-broadening signature, expected from an accreting disk around a
protostar, is
in order of 300~km s$^{-1}$, typically. With the energy resolution of 5~eV and
the energy determination accuracy of 0.5~eV, the detection of such a Doppler
signature is well within the realm of possibility. For diffused X-ray emission,
the same skill can be applied to
separate unidentified emission lines observed in CCD spectra from known thermal
emission lines as well, as their identification requires measurements of their
line center energies at less than a few eV uncertainty.

\smallskip
Astrophysicists have models to ascribe X-ray phenomena witnessed in those
stellar targets, but not enough data to test them with rigor. So we have
awaited the advent of the Soft X-ray Spectrometer on board {\it ASTRO-H}. In conjunction
with other instruments, especially the Hard X-ray Imagers, we will be able
to explore a new horizon to extend our understanding of stars and related phenomena.
\clearpage

%%%%%%%%%%%%%%%%%%%%%%%%%%%%%%%%%%%%%%%%%%%%%%%%%%%%%%%%%%%%
\section{Protostars} 
%\author{Prepared by Y. Tsuboi, H. Murakami, and Y. Maeda}
%%%%%%%%%%%%%%%%%%%%%%%%%%%%%%%%%%%%%%%%%%%%%%%%%%%%%%%%%%%

\subsection{Background and Previous Studies}

%A protostar is a phase of accretion 

Low-mass young stellar objects (YSOs) evolve from molecular cloud
cores through the protostar, Classical T Tauri (CTTS) and Weak-line T
Tauri (WTTS) phases to the main sequence.  Protostars are generally
associated with Class 0 and I spectra, with spectral energy
distributions (SEDs) which peak in the mm and mid- to far-IR bands,
respectively. Bipolar flows accompany this phase, indicating dynamic
gas accretion.  However, due to very high extinction in the optical
band, the vicinity of the {\it protostar itself} and the process of
accretion onto the central star had never been directly observed.

%CTTSs are in quasi-static
%contracting phase of the star itself, although they have still 
%circumstellar disks. They are associated with Class II spectrum, which peaks
%at near-infrared.  Finally, as the circumstellar disk disappear, YSOs
%evolve to WTTSs, associated with Class III spectrum (Shu et al. 1987,
%Andre and Montmerle 1994).

%With the {\it Einstein} satellite, TTSs (CTTSs and WTTSs) has been found
%to emit soft X-rays intensively, showing X-ray characteristics --
%moderate plasma temperatures, strong variability with occasional rapid
%flares -- consistent with an enhanced solar-type activity of magnetic
%field produced by the dynamo mechanism (Feigelson and DeCampli 1981,
%Montmerle et al. 1983).

%However, due to the very high extinction in optical
%band, the very vicinity of the *protostar itself* and the process of
%accretion onto the central star had never been directly observed.

This breakthrough was achieved in {\it hard} X-rays due to high
transparency of the ISM in this band. X-rays from protostars were
discovered with {\it ASCA} in $\rho$ Oph \citep{Koyama94} and detailed
analyses were performed by \cite{Kamata97}. They detected three Class
I protostars, which all have a spectrum characterized by heavy
absorption ($N_{\rm H} > 10^{22}$ cm$^{-2}$) and highly ionized iron K
shell lines. The spectra are well-fitted with optically thin thermal
plasma model with $kT >$ 1 keV. Who would have expected that the
central star in a molecular core has such high temperature gas, while
the molecular core has only tens of Kelvin? Flare-like time
variability was observed in one of the Class I protostars (EL 29).
Such variability is seen on the solar surface and implies magnetic
activity on/near the central star of the protostar. The time scale of
the flare and the temperature of the gas are greater than those in
solar flares, and the X-ray luminosity of the protostar is
$10^3$--$10^6$ times greater than solar flares. 

The Class I protostar WL 6 showed a sinusoidal light curve with a
period of about one day, and with constant temperature, which suggests
rotational modulation of the protostar \citep{Kamata97}. On the other
hand, the Class I protostar YLW 15A showed quasi-periodic X-ray flares
with about 20 hour intervals \citep{Tsuboi00}. The flares can be
interpreted as a star-disk interaction in which the star itself is
rotating rapidly with 1 day spin period \citep{Montmerle00}. The
period of 1 day is significantly shorter than the spin periods in the
other phases (about 7 days in Class II and 3 days in Class III). If
the stellar spin is truly 1 day, the central star might rotate with
nearly break-up velocity. The rotation period is undoubtedly a key
parameter for accretion onto protostars; it is a direct indicator for
the accumulated angular momentum of the star itself. These results
have demonstrated that we can derive the fundamental parameter
uniquely with hard X-ray band.

It was demonstrated that we can probe not only the protostar itself,
but also its close vicinity, using emission lines. \cite{Imanishi01}
discovered a fluorescent iron line at 6.4~keV as well as the highly
ionized iron line at 6.7~keV from one of the protostars, YLW16A, in
its flare phase. Similar spectra have been observed in type I Seyfert
galaxies. Based on this similarity, they interpret that the incident
X-rays emitted from the protostar itself irradiate a surrounding cold
material which then re-emits fluorescent lines. The large equivalent
width (100 eV) and not so large absorption column (5$\times10^{22}$
cm$^{-2}$) is comparable to what is observed in type I Seyfert
galaxies. If the circumstellar gas is spherically distributed around
the X-ray source, the equivalent width of the iron line is predicted
to be just $\sim$ 15 eV.  Hence, nonspherical geometry is required; a
larger amount of gas is present out of the line of sight,
i.e. reflection by a face-on disk is plausible
\citep{Sekimoto97}. Since the time lag between the flare onset and the
6.4 keV iron line appearance is shorter than 3 hours, the separation
between the star and the reflecting region is $\leq$20 AU,
consistent with a disk origin. After this discovery, neutral Fe
fluorescent lines were also detected from other several embedded
sources (\citealt{Tsujimoto04}, \citealt{Hamaguchi05}), including our
proposed target R CrA IRS 7.

\begin{figure}[ht]
\begin{center}
\includegraphics[width=4.8in,angle=0.0,clip]{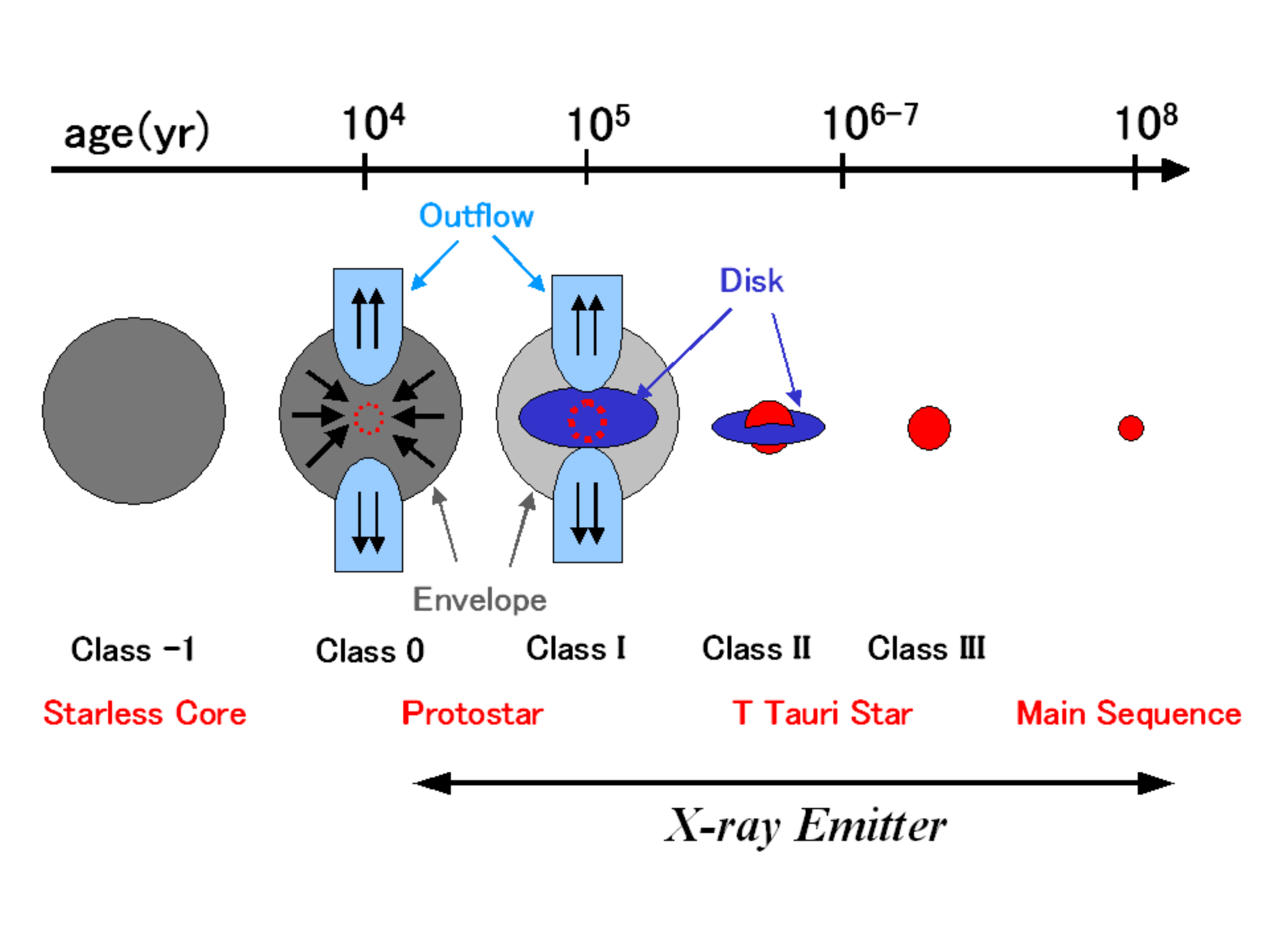}
\caption{Schematic view of star formation. Protostars are classified
  as Class 0 and I, which have SED peaks in the $mm$ and mid- to
  far-IR bands, respectively. Bipolar flows accompany these phases,
  indicating dynamic gas accretion.
\label{figure:YSO}}
\end{center}
\end{figure}

\subsection{Prospects \& Strategy}
With the above insights, we believe that we can probe the
structure and the dynamics of central stars in dense molecular cores
and those of the vicinities using the following methods.  Although
observations of radio masars can probe in to a distance of about
10 AU from the central protostar, the most inner part can only be
revealed with the high-resolution spectroscopic capability of
the SXS. 

\paragraph{(1) The spin and the radius of the central star of a protostar}
\ 

By monitoring the doppler-shifted He-like iron line (6.7 keV), we will
obtain a rotation period of the X-ray-active region possibly on the
protostar itself, directly.  If a protostar has a radius of $\sim 5
{\rm R}_{\odot}$ and a rotation period of 1 day, the maximum Doppler
velocity broadening due to this rotation are $\sim$350~km~s$^{-1}$
sin~$i$ and 8 eV sin~$i$ at 7 keV, respectively, where $i$ is the
inclination angle of the star's rotational axis.  The {\it ASTRO-H}
SXS, with its superior spectral resolution ($E / \Delta E$ = 1000) at
7 keV, is the only spectrometer capable of detecting this Doppler
broadening. This measurement can be used to infer the stellar radius:
\begin{equation}
  2 \pi R_* \cos \theta \sin i = v \cos \theta \sin i\, T\, ,
\end{equation}
where $\theta$ is the latitude of the active region and $i$ is the
inclination angle of the stellar rotation axis.  The spin velocity and
the stellar radius probe the angular momentum of the protostar itself,
which is crucial in understanding the contraction process of a star.

\paragraph{(2) The dynamics of the accreting matter}
\

The accretion disk of a protostar emits 6.4 keV photons from neutral
iron due to fluorescence by hard X-rays from the star itself.  By
measuring the Doppler shift and broadening of this line we will
characterize the dynamics of the inner part of the disk.  The
differences of orbital period and velocity between the disk and the
stellar surface, as characterized by the 6.7 keV, allow us to
understand the transfer of angular momentum from the disk to the star.

As with the 6.7 keV line, we can derive the radius of the 6.4 keV emitting
region. In this case, by assuming Keplerian motion, we can
also obtain a lower limit to the enclosed mass, i.e. the mass of the
protostar itself.

By making time-sliced spectra during a flare and measuring the neutral
Fe K equivalent width, as \cite{Imanishi01} did, we can construct a map of the structure of the
inner accretion disk.  Because of the high
sensitivity of the SXS in the Fe K band in comparison with {\it Chandra}, 
we can characterize this structure with higher resolution.
The structure of the inner part of the protostellar accretion disks is
of particular interest, since these disks will evolve into planetary
disks.

\paragraph{(3) Acceleration of the jets from protostars}
\

\cite{Pravdo01} detected jets from the protostar HH2 for the first
time in the X-ray band. The velocity of the bulk motion of the jets
was estimated as 200 km s$^{-1}$ from the plasma
temperature. \cite{Favata02}, \cite{Bally03}, \cite{Tsujimoto04}, and
other recent studies also found fast X-ray jets with a speed of
500--1000 km s$^{-1}$. Jets are expected to accompany X-ray flares,
which occur due to magnetic reconnection events (e.g.,
\citealt{Hayashi96}).  In fact, such phenomena have been observed on
the Solar surface. The {\it ASTRO-H} SXS can search for co-existence of
X-ray flares and jets from a protostar by looking for the Doppler
signature of the jet in emission lines contemporaneously with an
increase in count rate.

\paragraph{(4) Plasma temperature}
\ 

In past protostar surveys, it has been found that younger stars have
X-ray emission from higher temperature plasma. However, in significant
numbers of protostars, the temperatures are not well-determined, owing
to limited statistics, heavy circumstellar absorption, and the drop in
sensitivity of conventional X-ray telescopes above 10 keV.
Measurements of the line ratios between He-like Fe at 6.7 keV and the
H-like Fe at 6.9 keV allow us to infer plasma temperatures
independently of the X-ray continuum spectrum. The sensitivity of SXS
in the Fe K band will allow us to determine the temperature with
unparalleled precision.

\paragraph{(5) The turn-on age of stellar X-rays}
\

{\it ASCA} detected X-rays from embedded Class I protostars
\citep{Koyama94} and {\it Chandra} is now establishing the X-ray
properties of Class I. Seventy percent of Class I objects were
detected in a 100 ks {\it Chandra} observation of the Ophiuchi cloud
\citep{Imanishi01}, demonstrating that virtually all Class I YSOs emit
X-rays.  On the other hand, none of the 41 pre-stellar Class $-1$
cores in the Ophiuchi cloud were detected (Figure~\ref{figure:YSO}),
indicating that the protostar itself is essential for X-ray
production. These results indicate that X-rays turn on at some point
during the Class 0 phase, but the precise turn-on age is still an open
question.

Some Class 0 protostars have been tentatively detected as X-ray
sources, but whether these sources are bona-fide Class 0 objects is
still controversial: in the Log $M_{\rm{env}}$ -- Log $L_{\rm{bol}}$
diagram, they are all located in the region corresponding to the Class
I phase or the boundary region between the Class I and 0 phases
\citep{Getman07}. The difficulty in detecting X-ray emission from
bona-fide Class 0 protostars might be due to the presence of larger
absorbing columns than those in Class I protostars. The hard X-ray
band above 10 keV is even more transparent than the {\it Chandra}
band, and thus the {\it ASTRO-H} HXI is extremely sensitive to X-rays from
protostars embedded in thick clouds of with circumstellar absorbing
columns of $N_H \ge 10^{24}$ cm$^{-2}$.

\paragraph{(6) Probing of soft gamma-ray precursor of flares}
\ 

Solar flares show a precursor to thermal X-ray flares in the
non-thermal soft gamma-ray band \citep[e.g.,][] {Sakao92}. The
precursor is thought to be non-thermal bremsstrahlung emission which
occurs as a result of particle acceleration during magnetic
reconnection events. No apparent precursor has yet been observed from
stars other than the Sun.  The HXI will be very sensitive to this
precursor emission, while the SXI and SXS will detect K-shell lines
from highly ionized Fe emitted in the thermal part of flares. The
combination of these {\it ASTRO-H} instruments is a strong tool in
characterizing the flaring process in protostars.

\subsection{Targets \& Feasibility}
There are two nearby young stellar clusters that include protostars
bright enough to do spectroscopy with the {\it ASTRO-H} SXS
calorimeter: $\rho$ Oph ($d \sim 165$ pc), and R CrA ($d \sim 130$
pc).  In the 2.9$'\times2.9'$ field of view of the SXS, four Class I
protostars can be included in an observation of part of the $\rho$ Oph
region, and three Class I protostars and two Class 0 protostars can be
included in an observation of the R CrA region.

\begin{figure}[ht]
\begin{center}
\includegraphics[width=0.8\textwidth,clip,angle=0]{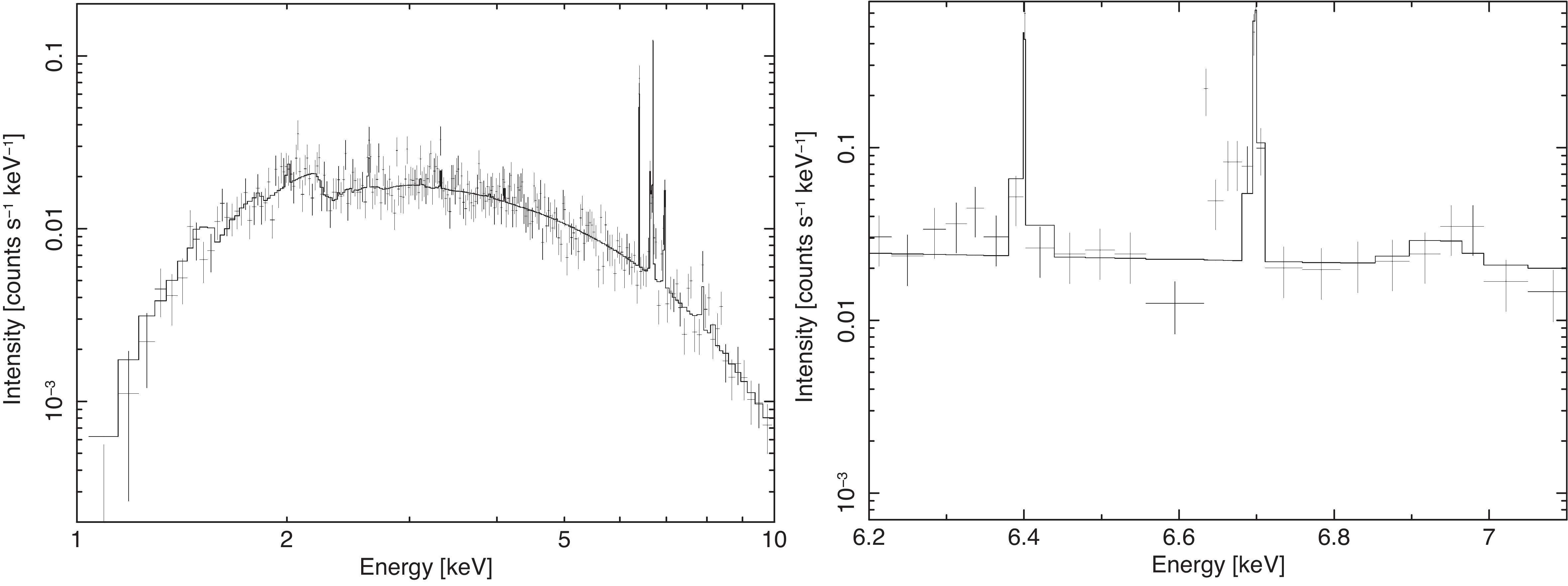}
\vspace{-0.3cm}
\caption {Simulation of the {\it ASTRO-H}/SXS spectrum of a
  protostar. Left panel: View of the broadband spectrum, assuming an
  X-ray luminosity of $1\times10^{31}$~ergs~s$^{-1}$ and exposure time
  of 80 ks. Right: detailed view of the iron K$_\alpha$ band. An X-ray
  luminosity of $4\times10^{31}$~ergs~s$^{-1}$ and 10 ks exposure time
  are assumed ($\sim$ 1/4 day with 50\% observing efficiency).}
\label{figure:simulation}
\end{center}
\end{figure}

To verify the rotational period of a given system to be $\sim$ 1 day,
we should compare the X-ray spectra of each 20~ksec ($\sim$ 1/4
day). The observing efficiency of {\it ASTRO-H} will allow us to achieve
10~ksec effective exposure.
Assuming $L = 10^{31}$~ergs~s$^{-1}$, $kT = 6$~keV, $N_{\rm
  H}$~=~3~$\times 10^{22}$~cm$^{-2}$, and elemental abundances of 0.3
solar, we will be able to constrain the Doppler shift of He-like iron
lines with a 1-$\sigma$ uncertainity of 1.5 eV,
Thus, the rotational speed of the protostar can be determined with
100~km~s$^{-1}$ precision. If we detect the 6.4-keV neutral iron line
with an equivalent width of 100~eV, the uncertainity on the Doppler
shift and broadening will be comparable to that for the 6.7-keV line.

\begin{figure}[ht]
\begin{center}
\vspace*{-0.5cm}
\includegraphics[height=7cm,clip]{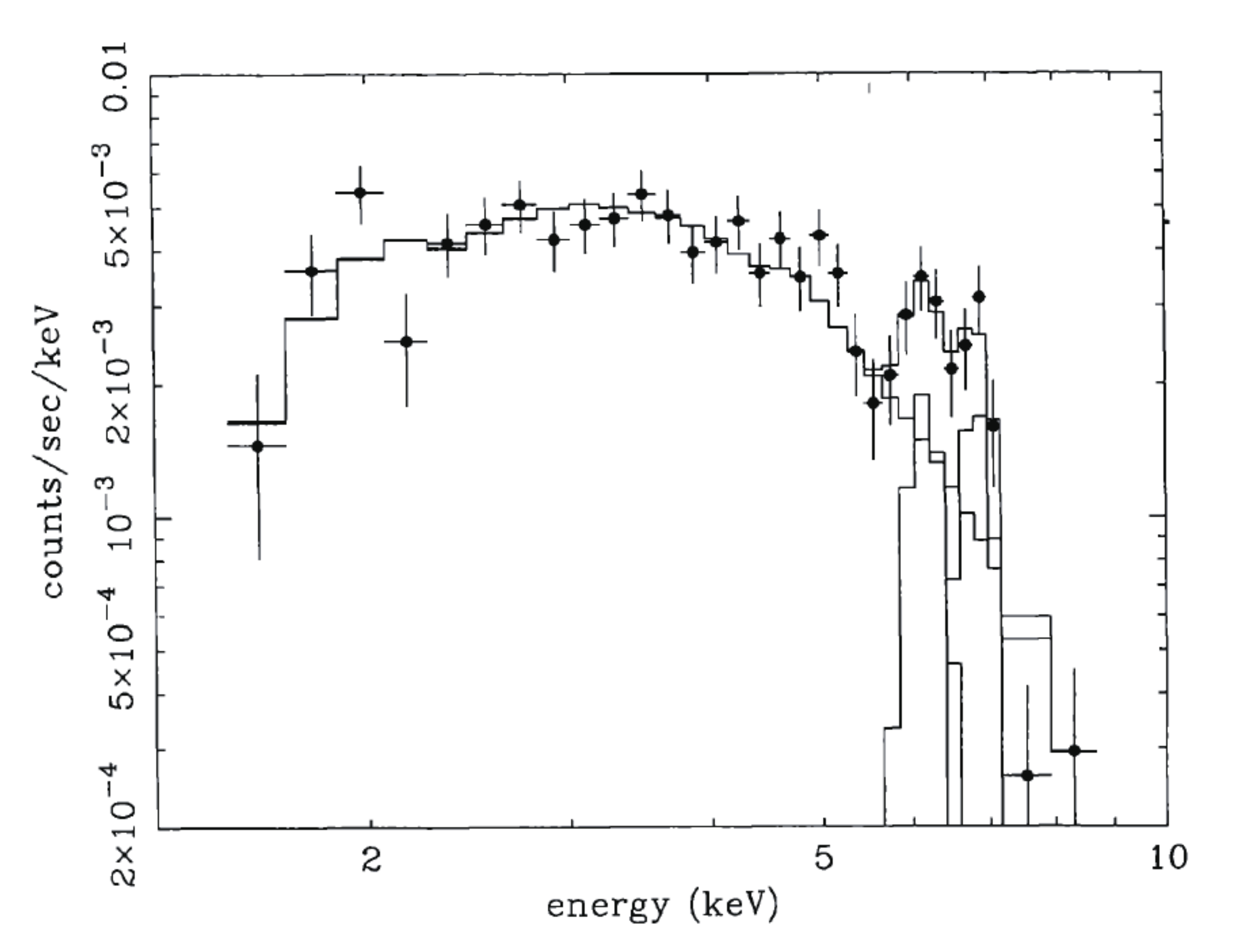}
\caption{
	Spectrum observed in a flare of IRS7, obtained with 
	{\it ASCA} \citep{Koyama96}.
}
\label{figure:RCrA_ASCA_spec}
\end{center}
\end{figure}

Note that one of the Class 0 candidates (X$_{\rm E}$, or Source 8 in
\citealt{Feigelson98}) flared during an {\it XMM-Newton} observation in March 2003
\citep{Hamaguchi05}. During the flare, a strong Fe K fluorescent line
at 6.4 keV was detected with an equivalent width of about 800 eV. Also
note that in the {\it ASCA} observation of this region
\citep{Koyama96}, striking results were obtained: during a flare on
IRS 7, two line features appeared with central energy of 6.81 keV and
6.12 keV (Figure~\ref{figure:RCrA_ASCA_spec}). If we assume that the
6.12 keV feature is a Doppler-shifted 6.4 keV fluorescent line from
neutral iron, the inferred infall speed of the cold material would
exceed 10$^4$ km s$^{-1}$, which is orders of magnitude larger than
expected for low-mass stars. If we catch such an enigmatic flare with
the SXS, spectra with clearly resolved lines will be obtained, and the
dynamics of the high-speed plasma will be easily resolved.

\clearpage

%%%%%%%%%%%%%%%%%%%%%%%%%%%%%%%%%%%%%%%%%%%%%%%%%%%%%%%%%%%%
\section{Flare Physics in Low-mass Stars}\label{s1}
%%%%%%%%%%%%%%%%%%%%%%%%%%%%%%%%%%%%%%%%%%%%%%%%%%%%%%%%%%%

\subsection{Background and Previous Studies}\label{s1-1}
Since the {\it Chandra} and {\it XMM-Newton} observatories opened an era of X-ray
high-resolution spectroscopy a decade ago, low-mass stars are among
the most observed classes of sources by their grating
instruments. This is quite reasonable considering that they exhibit
diagnostically rich line-dominated spectra and are observationally
easy targets. Low-mass stars are active in the X-ray band and hundreds
of objects can be found in our proximity which have well-determined
stellar parameters and distances, and intense X-ray flux sufficient
for high-resolution spectroscopy.  They are distributed in Galactic
longitude and latitude, thus are visible even when other Galactic
targets are not.

The X-ray emission from low-mass stars have been interpreted by
extrapolating from the Sun, in which the combination of differential
rotation and convection gives rise to surface magnetic fields, with
stored energy released in magnetic reconnection events, resulting in
the production of high-temperature plasma. However, not all low-mass
stars operate identically to the Sun. The Sun is a single star,
whereas most X-ray active low-mass stars are binaries. The Sun has a
rotation period of $\sim$30 days, while other low-mass stars rotate
much more rapidly, e.g. BO Mic, with a rotation period of
0.4~days. The Sun hosts a planet system with two gas giants (Jupiter
and Saturn) far away from itself, but an increasing number of stars
are now known to host gas giants very close to the primary star. All
these differences make X-ray emission from low-mass stars different from
those of the Sun both quantitatively and qualitatively.

A typical X-ray spectrum obtained in an observation of a low-mass star
is a mixture of soft, stable coronal emission, together with
occasional hard and transient flare emission. The flares are caused by
magnetic reconnection events, while the corona is an integrated energy
reservoir. While both {\it XMM-Newton} and {\it Chandra} were
well-suited to study the former, {\it ASTRO-H} is best suited for the
latter.  Although HETG has sensitivity at $>$2~keV, the effective
area is so small that it has not been possible to extract meaningful
constraints from Fe K line profiles during stellar flares. Although
HXI will have sensitivity similar to {\it NuSTAR}, simultaneous observation
by a high-resolution soft X-ray spectrometer in the soft-band and a
sensitive hard-band detector, which is a requisite in flare
observations, will only be possible with {\it ASTRO-H}. It is expected that
{\it ASTRO-H} will yield fruitful results in stellar flare physics, and we
discuss selected topics in this field below.

\subsection{Prospects \& Strategy}\label{s1-2}
\subsubsection{Dynamic movement of materials during Flares}\label{s1-2-1}

Recent solar X-ray observations have revealed dynamic motion during
flares. The energy released in magnetic reconnection events is given
to charged particles. While upward-moving particles give rise to a
coronal mass ejection, downward-moving particles evaporate cold
material in the chromosphere, which is trapped as hot plasma in flare
loops. Such movements are routinely captured in spatially resolved
movies. The movement of coronal material is traced e.g. by
Doppler-shifted Fe {\sc XV} emission 
%by 100-150~km~s$^{-1}$
\citep{asai08}.

Although we cannot obtain spatially resolved observations of stellar
flares, there are hints in the integrated spectrum which suggest that
such phenomena take place on a much larger scale.  For example, Algol
is a typical X-ray emitting stellar system, which is comprised of an
X-ray dim B star and an X-ray active K star. This source exhibits such
frequent X-ray flares that almost all pointed X-ray observations
captured a flare \citep{favata99,yang09}.

During a Beppo-SAX observation \citep{favata99} of Algol, two
interesting features were found. First, the elemental abundances
inferred from the X-ray spectrum obtained during a flare changed
compared to the pre-flare spectrum, and in particular the Fe abundance
increased by a factor of a few.  This is interpreted as evaporation of
Fe in the chromosphere into the corona. Second, photoelectric
absorption of the X-ray emitting region increased during the flare,
which is interpreted as absorption through coronal mass ejection (CME)
material ejected by the flare. These interpretations need to be tested
with direct measurements of the motion of the X-ray emitting
material. The expected Doppler velocity of $\sim$200~km~s$^{-1}$ will
be easily distinguished by the SXS.

\subsubsection{Particle Acceleration in Flares}\label{s1-2-2}

\begin{figure}[hbtp]
 \begin{center}
 \includegraphics[width=0.7\textwidth]{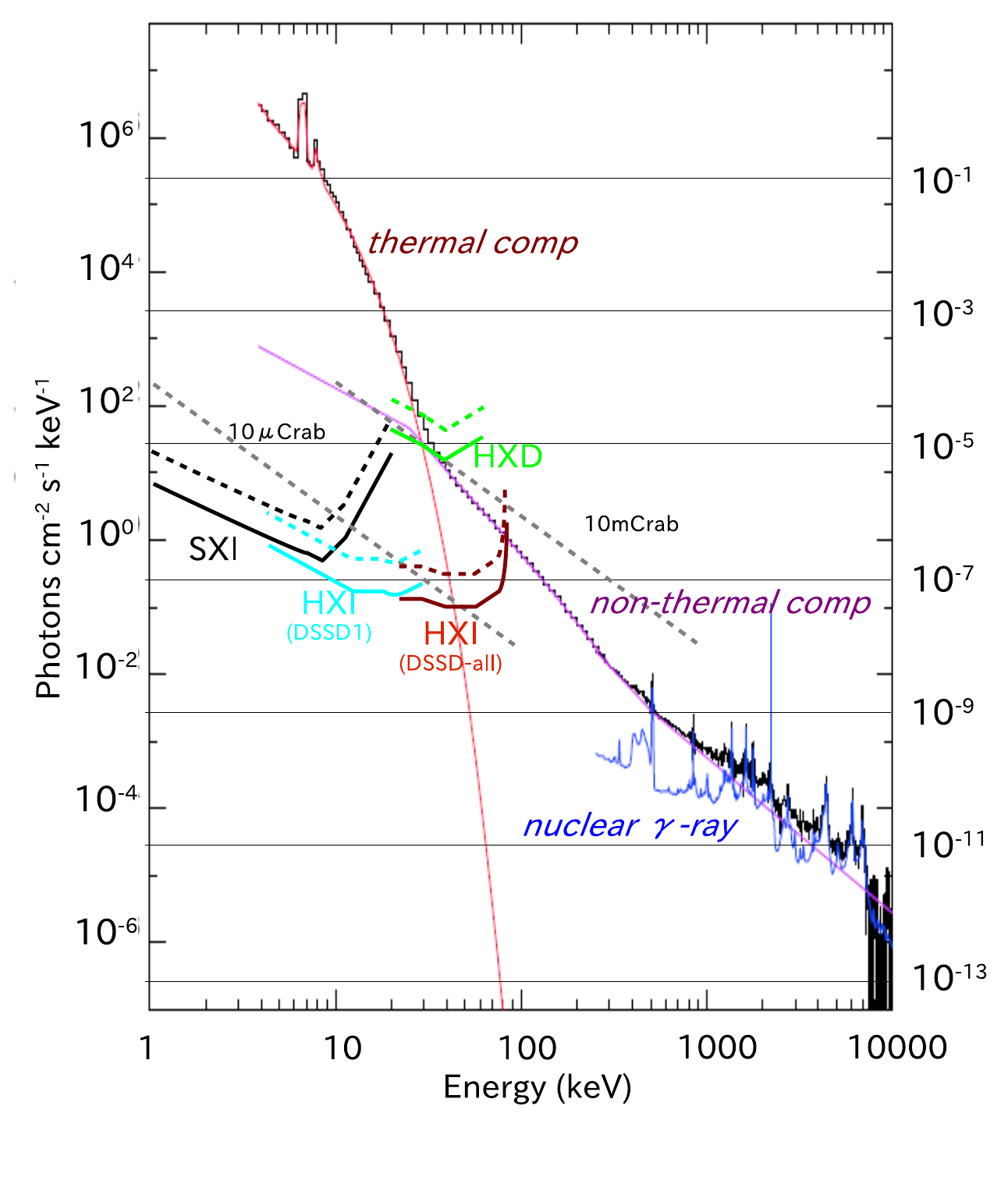} 
  \caption{Spectrum of an X4.8 class Solar flare \citep{lin11}, with
    the righthand $y$-axis scaled to the luminosity of a giant stellar
    flare observed in UV Cet \citep{negoro12}. The flare consists of
    three major spectral components (thermal, non-thermal, nuclear
    $\gamma$-ray lines). The SXI, HXI, and {\it Suzaku}/HXD sensitivity for
    100~ks observations (and for 1/10 flux) is shown with solid
    (dashed) curves.}
  \label{f02}
 \end{center}
\end{figure}

The RHESSI observatory has made revolutionary progress in
understanding high-energy phenomena in flares.  We expect that {\it ASTRO-H}
will make similar progress in understading stellar flares in the
10--50~keV band. Figure~\ref{f02} shows an X4.8 class solar flare
observed by the RHESSI observatory \citep{lin11}. The emission is
comprised of a thermal component below 30~keV, non-thermal
components from 30--1000~keV and nuclear $\gamma$-rays above 1~MeV.

The sudden release of magnetic energy in a flare gives rise to
accelerated charged particles. Downward-moving particles collide with
dense material in the atmosphere, where they emit non-thermal
bremsstrahlung as well as nuclear $\gamma$-rays. The released energy
is thermalized and the flare loop filled with thermal plasma.

The detection of nuclear $\gamma$-ray emission from stars is still far
below the sensitivity of contemporary technology, but the non-thermal
emission is within reach. \citet{osten07} reported detection with the
{\it Swift} BAT of hard X-ray emission extending to 200~keV during a giant
flare from II Peg. The spectrum was statistically inconclusive as to
whether it is due to non-thermal or high-temperature thermal plasma,
but it certainly illustrated that $>$10~keV emission does exist in
stellar flares like the Sun.

We expect that {\it ASTRO-H} will make two major advances in this
field. First, it will detect $>$10~keV X-rays from a number of
low-mass stars, greatly expanding the sample over UV Ceti, the only
one known to date. In Figure~\ref{f02}, we show the spectrum of a
Solar X4.8 class flare, with the righthand $y$-axis scaled to the
luminosity observed by MAXI from a giant flare on UV Cet
\citep{negoro12}, and with a comparison to the expected sensitivity by
the {\it ASTRO-H} instruments as well as {\it Suzaku} HXD. The flux of the
non-thermal emission is just below the HXD sensitivity but far above
that HXI sensitivity, demonstrating that we are standing just at the
edge of this discovery space.

{\it ASTRO-H} will also conclusively determine whether the $>$10~keV
emission in stellar flares is non-thermal, and will measure the
low-energy cutoff the accelerated particles for the brightest
cases. The low-energy cut-off is a key parameter which remains to be
constrained even in solar flares. The photon index of the non-thermal
emission is large (3--6; see Fig~\ref{f02}), so the estimate of the
number of accelerated electrons (the electric current from the corona
to the photosphere) can vary significantly with a slight change in the
low-energy cutoff. If the cutoff energy is 20~keV in a typical solar
flare spectrum, the number of accelerated electrons is absurdly large;
all the electrons in a volume larger than a solar active region need
to be accelerated, resulting in a current of $\approx$10$^{18}$~A.
There is thus speculation that the cut-off energy in the non-thermal
emission is significantly higher than this.

In order to make these measurements, we need to have a combination of
(i) a $<10$~keV spectrometer with sufficient energy resolution to
construct the differential emission measure precisely at the
high-temperature end using the Fe and Ni K emission lines and (ii) a
sensitive $>$10~keV detector, with both detectors operating
simultaneously to observe the same flare. This combination has never
been employed even in solar observations.

\subsubsection{Star-Planet Interaction}\label{s1-2-3}

Since it has became increasingly clear that close-in giant planets are
not so uncommon, an enhancment of flaring activity on the host star
due to star-planet interactions (SPIs) has been a matter of lively
debate \citep{cuntz00, rubenstein00}. SPIs are expected to include not
only tidal interactions one but also magnetic interactions that might
increase coronal activities and thus could be investigated through
X-rays \citep[e.g.,][]{lanza09}. If this is the case, an X-ray study
of such flare events would be a useful probe of magnetic field
strengths and configurations of extra-solar giant planet
systems. Although X-ray investigations of individual systems and
statistical samples have been both carried out \citep[][and references
  therein]{miller12}, the results have been mixed, and high-energy
activity induced by SPIs is still controversial. High quality data to
derive more reliable physical parameters, such as plasma density and
temperature, are in strong demand.

Since the time duration of X-ray flares observed so far from giant
planet systems is typically short ($\sim$10~ks in the case of
HD~189733; \citealt{pillitteri10, pillitteri11}) the large effective
area and high energy resolution provided by the SXS are required to
determine the plasma density and temperature of the flares. The SXS
has four times more effective area than the {\it XMM-Newton} and 
{\it Chandra} gratings near 1~keV in the Ne~K band, which is especially
useful to constrain the coronal density of the host star that, which
in the case of HD~189733 has been suggested to be unusually dense
\citep{pillitteri11}.

\subsection{Targets \& Feasibility}\label{s1-3}
\subsubsection{Flare Stars}\label{s1-3-1}

Sufficiently bright sources are needed to make use of the
high-resolution spectroscopic capability of the SXS, which has
2$^{14}$ spectral bins. In order to have a few counts per bin within a
reasonable exposure time, we require sources with brightness of at
least a few mCrab on average. Fortunately, we have hundreds of sources
\citep{makarov03} to choose from. Some of these sources exhibit flares
so frequently that it almost certain that we will detect a flare or two in
a $\approx$100~ks exposure.

\begin{figure}[hbtp]
 \begin{center}
 \includegraphics[width=0.7\textwidth,angle=0]{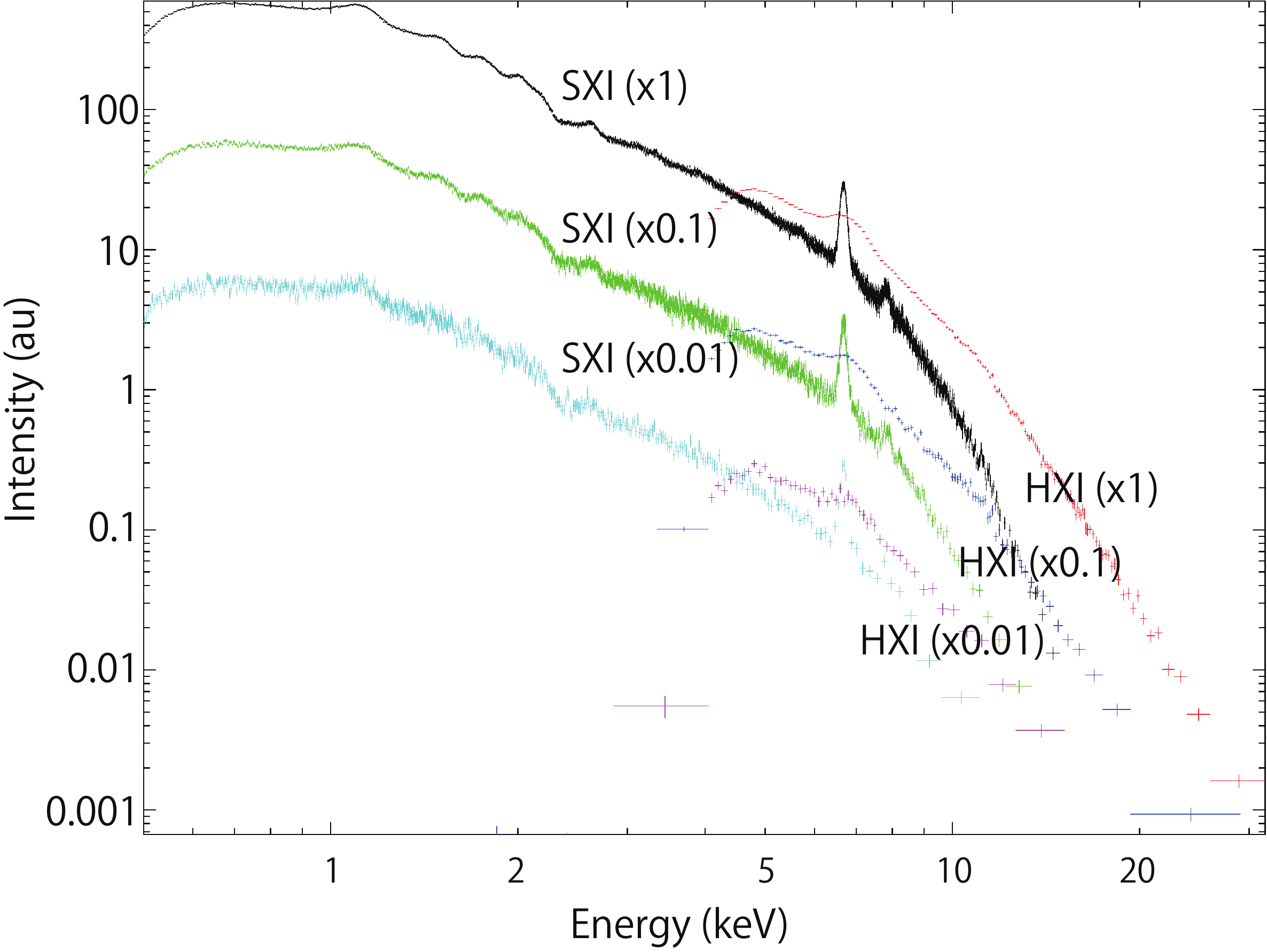}
  \caption{Simulated SXI$+$HXI spectra assuming an observed solar
    flare spectrum \citep{lin11} and a flux of $\times$1, $\times$0.1,
    and $\times$0.01 times of that observed in a large flare from UV
    Cet \citep{negoro12} for 10~ks. Pile-up in the SXI detector is not
    considered.}
\label{f01}
 \end{center}
\end{figure}

\begin{figure}[hbtp]
 \begin{center}
 \includegraphics[width=0.45\textwidth,angle=270]{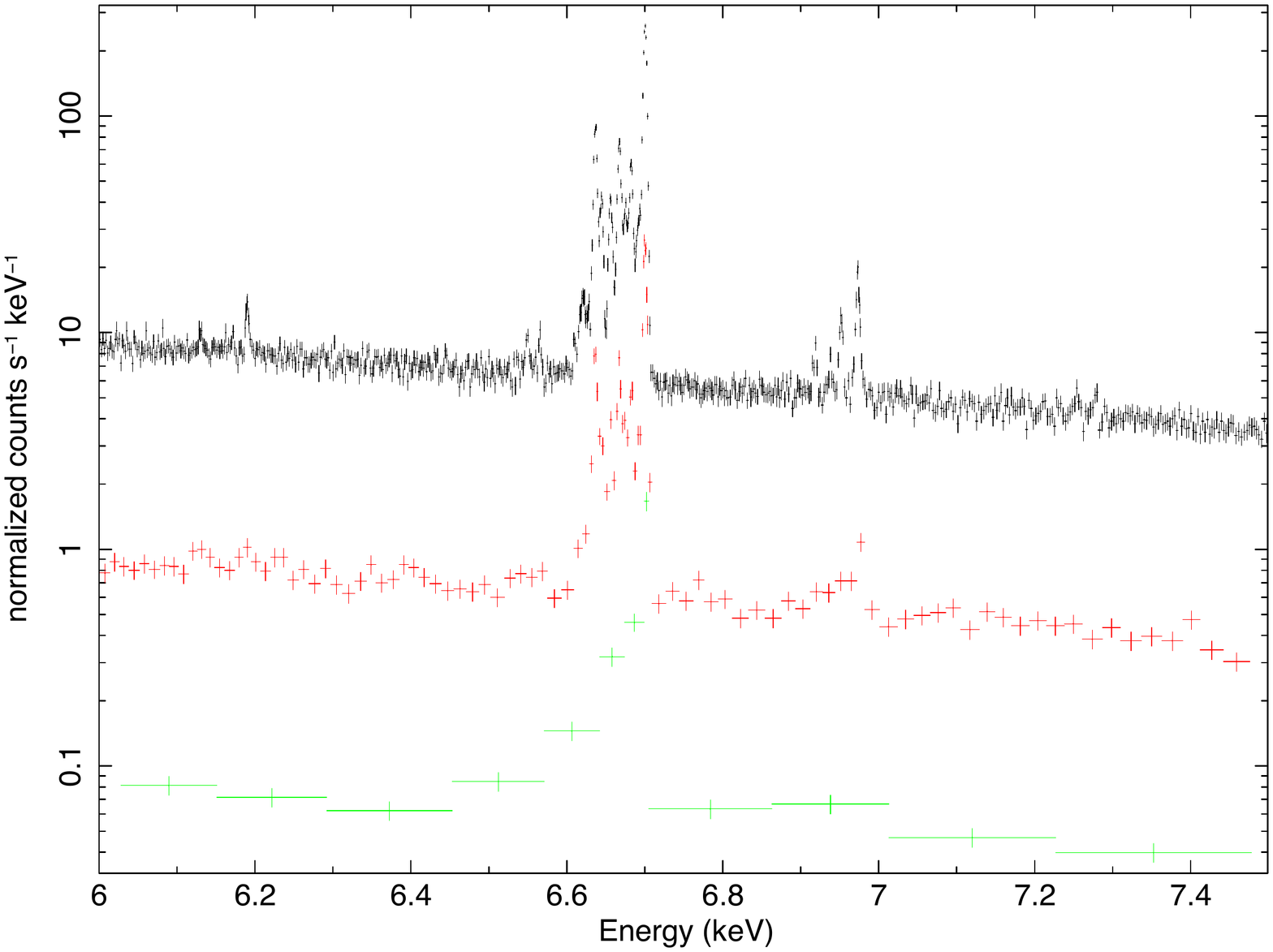} 
 \caption{Simulated SXS spectra around Fe K emission assuming an
   observed solar flare spectrum \citep{lin11} and a flux of
   $\times$1, $\times$0.1, and $\times$0.01 times of that observed in
   a large flare from UV Cet \citep{negoro12} for 10~ks.}
 \label{f03}
 \end{center}
\end{figure}

Nearby ($<$50~pc) bright sources that exhibit frequent flares are
prospective candidates (e.g., VY Ari, UX Ari, V711 Tau, $\sigma$ Gem,
BO Mic, AR Lac, and II Peg). In figure~\ref{f01}, we show simulated
SXI$+$HXI spectra assuming spectral distribution based on observations
of a Solar flare, and with different fluxes.  In the brightest case,
the low energy cutoff in the power law at 20~keV is seen in the HXI
spectrum. Figure~\ref{f03} shows the simulated SXS spectra in the Fe K
band for the same scenarios as in figure~\ref{f01}. A Doppler velocity
of $\sim$100~km~s$^{-1}$ is statistically distinguishable in the
brightest two cases.

\subsubsection{Planet-hosting Stars}\label{s1-3-2}

\begin{figure}[htbp]
 \centering
 \includegraphics[width=\textwidth]{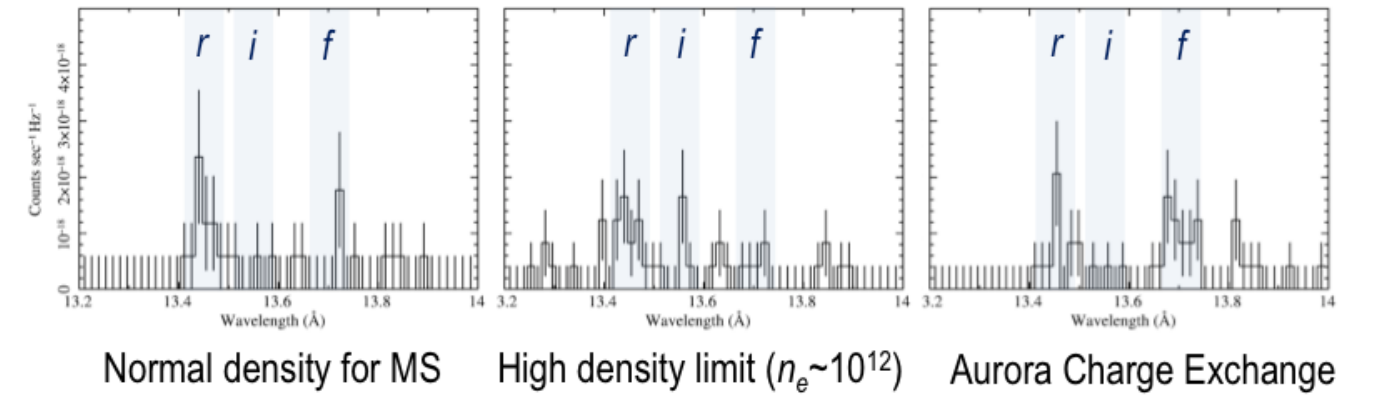} 
 \caption[]{SXS simulations of a flare on HD 189733 in three
   different scenarios. Left: normal low density scenario for
   main-sequence stars. Center: high density scenario
   ($n_{e}\sim10^{12}$). Right: auroral charge-exchange scenario.}
 \label{fig:HD189733}
\end{figure}

HD 189733 is a primary target in this category, since this system
showed an X-ray flare after secondary eclipse in two different
occasions and thus has been suggested to possess a systematic SPI when
the planet passes close to active regions on the host star
\citep{pillitteri11}. The increased X-ray flux due to the flare mainly
come in the Ne K band near 1 keV. Figure~\ref{fig:HD189733} shows SXS
simulations of Ne lines during the flare from HD 189733 in three
different cases, using parameters taken from a previously observed
flare \citep{pillitteri11}. Even with the large effective area of SXS,
the statistical quality of the data is not sufficient to strongly
constrain the plasma density. However, it still allows us to
distinguish the low and high density limiting cases. It will also
allow us to distinguish coronal emission from auroral charge-exchange
emission, as has been observed in the polar regions of Jupiter
\citep[e.g.,][]{hui09}.

\clearpage

%%%%%%%%%%%%%%%%%%%%%%%%%%%%%%%%%%%%%%%%%%%%%%%%%%%%%%%%%%%%
\section{Accretion Processes in Classical T Tauri Stars}
%%%%%%%%%%%%%%%%%%%%%%%%%%%%%%%%%%%%%%%%%%%%%%%%%%%%%%%%%%%

%\author{Prepared by M. Audard and K. Ishibashi}

\subsection{Background and Previous Studies}

The origin of the X-ray emission in young, accreting stars remains
unclear. It has been known for decades that young stars display strong
X-ray fluxes comparable to or higher than those in magnetically active
stars (e.g., \citealt{feigelson99}). High temperatures of several to
several tens of millions of Kelvin and flaring both point to enhanced
magnetic activity. No significant differences are observed between
weakly accreting and strongly accreting T Tau stars, suggesting that
accretion plays little if any role in the X-ray emission of young
stars, despite the strong interplay expected between the accretion
disk, infalling matter and the stellar magnetosphere.

This picture significantly changed with the advent of gratings onboard
{\it Chandra} and {\it XMM-Newton}.  \citet{kastner02} obtained
the first X-ray grating spectrum of TW Hya, a moderately accreting
classical T~Tauri star (CTTS), and unexpectedly observed very low
forbidden-to-intercombination lines ratios in the O~{\sc vii} and
Ne~{\sc ix} He-like triplets, indicating high electron densities of
the order of $10^{12}-10^{13}$ cm$^{-3}$. The effect of UV
photoexcitation on the observed line ratio was deemed negligible. In
addition, the X-ray spectrum was consistent with a quasi-isothermal
plasma of about 3 MK. Taken together, \citet{kastner02} concluded that
the X-ray emission in TW Hya was due to accretion, a result further
supported by observations with {\it XMM-Newton} \citep{stelzer04}
and with deeper {\it Chandra} observations
\citep{raassen09,brickhouse10}.

\begin{figure}[!b]
\includegraphics[width=.475\textwidth]{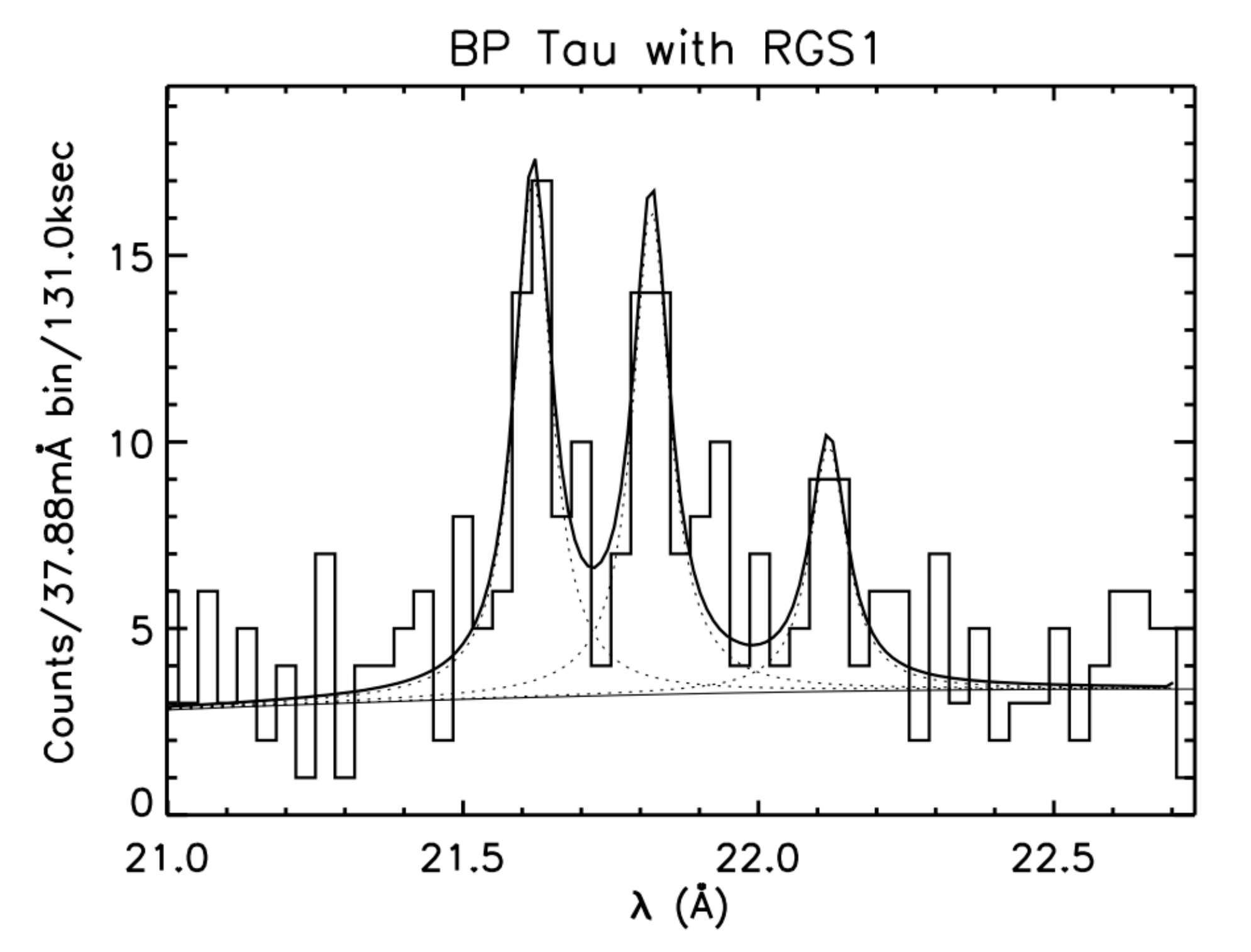}
\includegraphics[width=.475\textwidth]{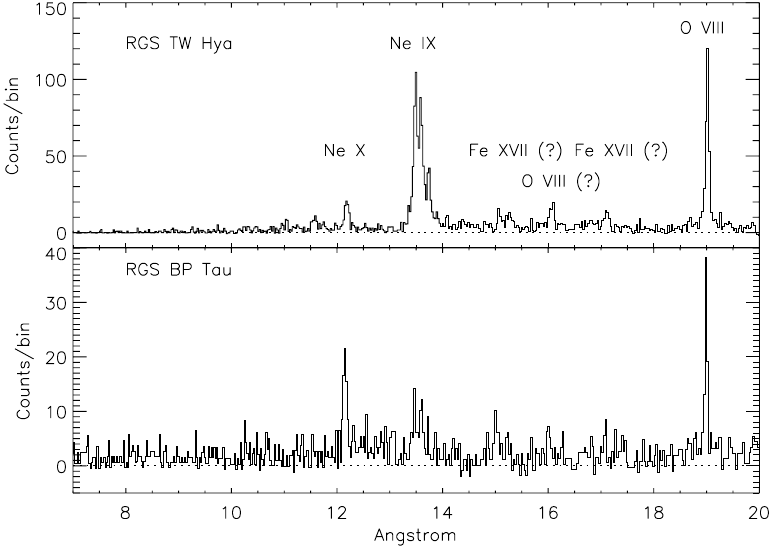}
\caption{{\it XMM-Newton} RGS1 spectrum of BP Tau; the left panel
  shows the O~{\sc vii} triplet, and the right bottom panel shows the
  Ne~{\sc ix} triplet, with the Ne~{\sc ix} triplet of TW Hya shown in
  the right top panel for comparison \citep{schmitt05}.  The exposure
  is 130~ks, showing the effective area limitations currently faced by
  grating spectroscopy.}
\label{fig:bptau_xmm}
\end{figure}

The TW Hya results triggered a set of grating observations of
CTTS. High electron densities were, indeed, found in the X-ray spectra
of several CTTS \citep[e.g.,][]{schmitt05, guenther06, robrade07,
  argiroffi07, argiroffi11, argiroffi12, huenemoerder07}.  Despite the
significant advancement made by grating spectroscopy, not many targets
are bright enough for grating observations, and the achieved
signal-to-noise ratios in typical exposures of the order of 100 ksec
are very low (see Fig.~\ref{fig:bptau_xmm}). High efficiency
non-dispersive spectroscopy capable of resolving He-like triplets
would be highly benefical to the understanding of the impact of
accretion on the X-ray emission in young stars; of particular
importance are the Ne~{\sc ix}, Mg~{\sc xi}, and Si~{\sc xiii}
triplets, where critical electron densities range from about $10^{11}$
to $10^{13}$ cm$^{-3}$, probing the range of densities expected in
accretion shocks.

Accreting CTTS further display a soft X-ray excess from plasma at low
temperature \citep{telleschi07,robrade07,guedel07}.  Such plasma,
mostly detected in the O~{\sc vii} lines, is difficult to detect using
CCD spectroscopy but can easily be revealed with high-resolution X-ray
spectra when the absorbing column density is not too high. A soft
X-ray excess can coexist with the hot plasma observed in the vast
majority of young stars, which is due to scaled-up solar-like magnetic
activity. The origin of such an excess is unclear, but it could be due
in part to accretion onto the stellar photosphere. It is, therefore,
crucial to understand how the X-ray emission in young stars, in
particular those accreting matter actively, can be influenced by the
accretion process.

While CTTS accrete matter in a somewhat "stable" fashion, though with
some variability, a handful of accreting young stars display powerful
eruptive events with flux increases in the optical regime of a few
magnitudes. Two classes have emerged: FUors, which display outbursts
of 4 magnitudes and more, last several decades, and therefore show a
low recurrence rate; and EXors (named after the prototype EX Lup),
which in contrast show somewhat smaller outbursts ($\Delta V=2-3$~mag)
on much shorter timescales, ranging from a few months up to a few
years, and which may occur repeatedly \citep[see review by][]{hartmann97}.
Such outbursts are believed to originate during
a rapid increase of the disk accretion rate from values of
$10^{-7}$~$M_\odot$~yr$^{-1}$ to $10^{-4}$~$M_\odot$~yr$^{-1}$ over a
short period of time. The limited number of eruptive young stars and
the long recurrence time (especially for FUor-type objects) make it
difficult to test models. It is therefore crucial to study the
evolution of outbursts in as much detail as possible.

\begin{figure}
\includegraphics[width=.45\textwidth]{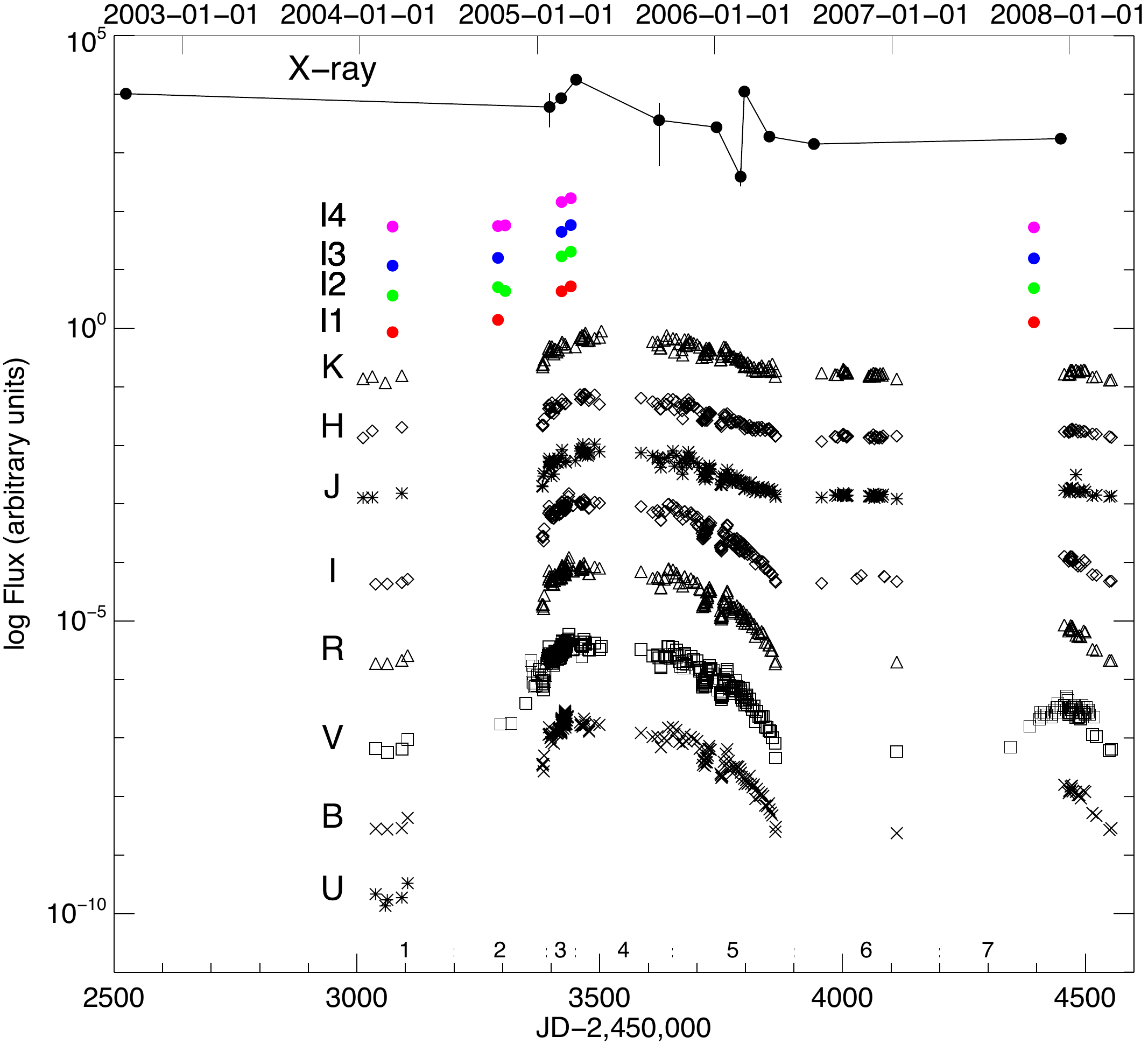}\hfill
\includegraphics[width=.50\textwidth]{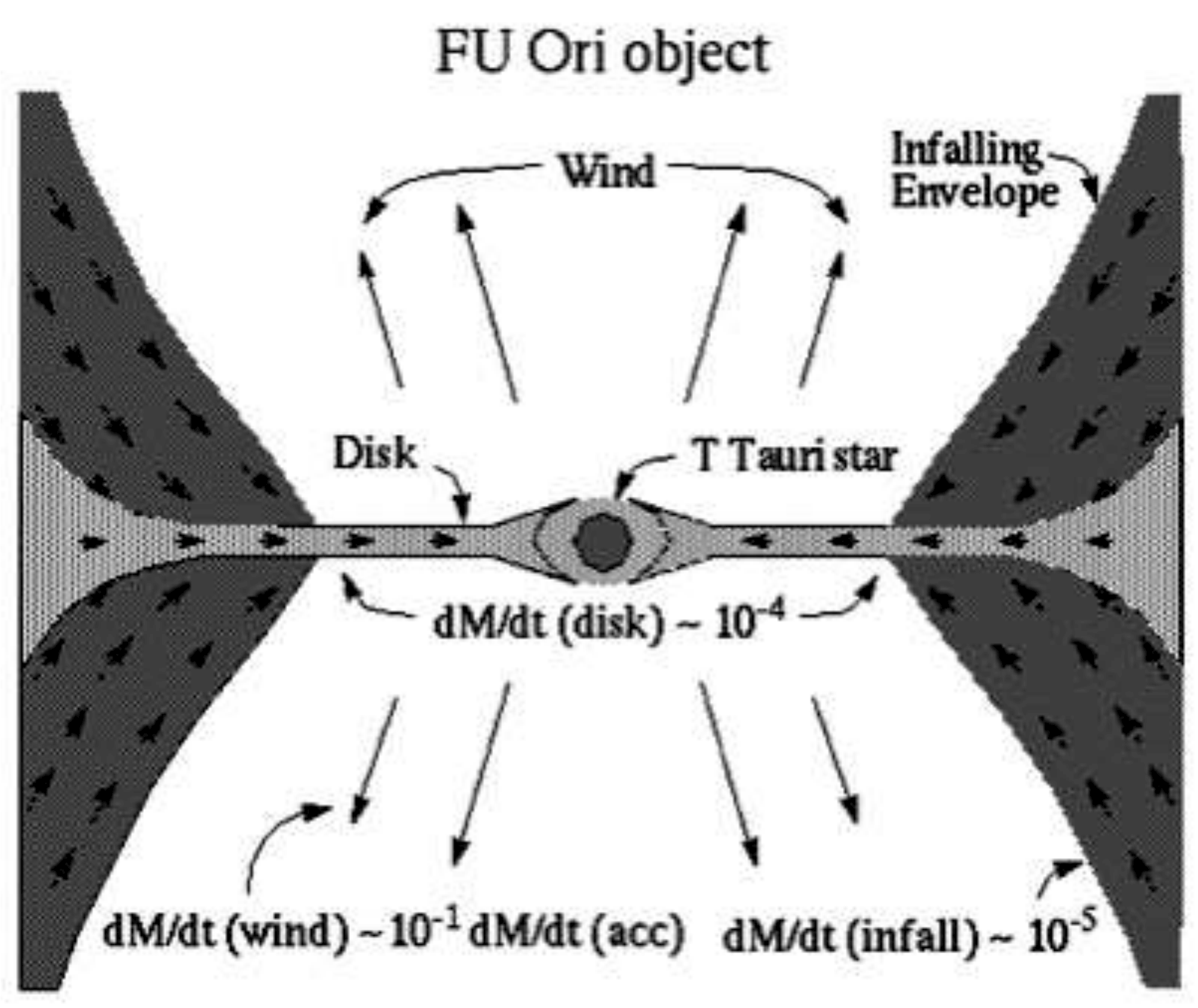}
\caption{{\it Left:} X-ray, optical, and infrared light curves of
  V1118 Ori before, during, and after its outburst \citep{audard10}.
  {\it Right: } Schema of an FU Ori object, showing the strong interplay
  between the accretion disk and the stellar magnetosphere
  \citep{hartmann97}.}
\label{fig:outbursting}
\end{figure}

In the high-energy domain, eruptive stars have gained significant
interest in the past few years.  Initially triggered by the outburst
of V1647 Ori, a young nearby low-mass star that erupted and
illuminated McNeil's nebula, multi-wavelength observations monitored
the outburst of other pre-main sequence accretion events and showed
that X-ray variability closely tracks optical and infrared variability
\citep[Fig.~\ref{fig:outbursting};][]{kastner04, kastner06, audard05,
  audard10, grosso05, grosso10, lorenzetti06, teets11, teets12,
  hamaguchi12}.  Although the exact behavior differs somewhat from
star to star, the overall interpretation is that accretion
significantly modified the magnetospheric configuration of the
star-disk systems during and after the events \citep{audard05,
  audard10, kastner06}. Nevertheless, recent observations suggest that
the star-disk magnetic field configurations may persist through
multiple outbursts \citep{hamaguchi12, siciliaaguilar12}. In contrast,
the X-ray spectra of FUors show hard emission \citep{skinner06,
  stelzer09}, with likely contamination by nearby nearby low-mass
companions \citep{skinner10}. While they demonstrate the strong
interplay between the accretion disk and the stellar magnetosphere,
the X-ray spectra are strongly flux-limited and studies have focussed
only on CCD spectroscopy.  No measurements of He-like triplet
diagnostics have yet been possible. The SXS will undoubtedly
contribute in an unprecedented fashion in the understanding of
accretion events in young stars.

\subsection{Prospects \& Strategy}

{\it ASTRO-H} and the SXS in particular will determine electron densities in
many young accreting stars thanks to its sensitivity and its
non-dispersive energy resolution.

\begin{figure}
\includegraphics[width=.44\hsize]{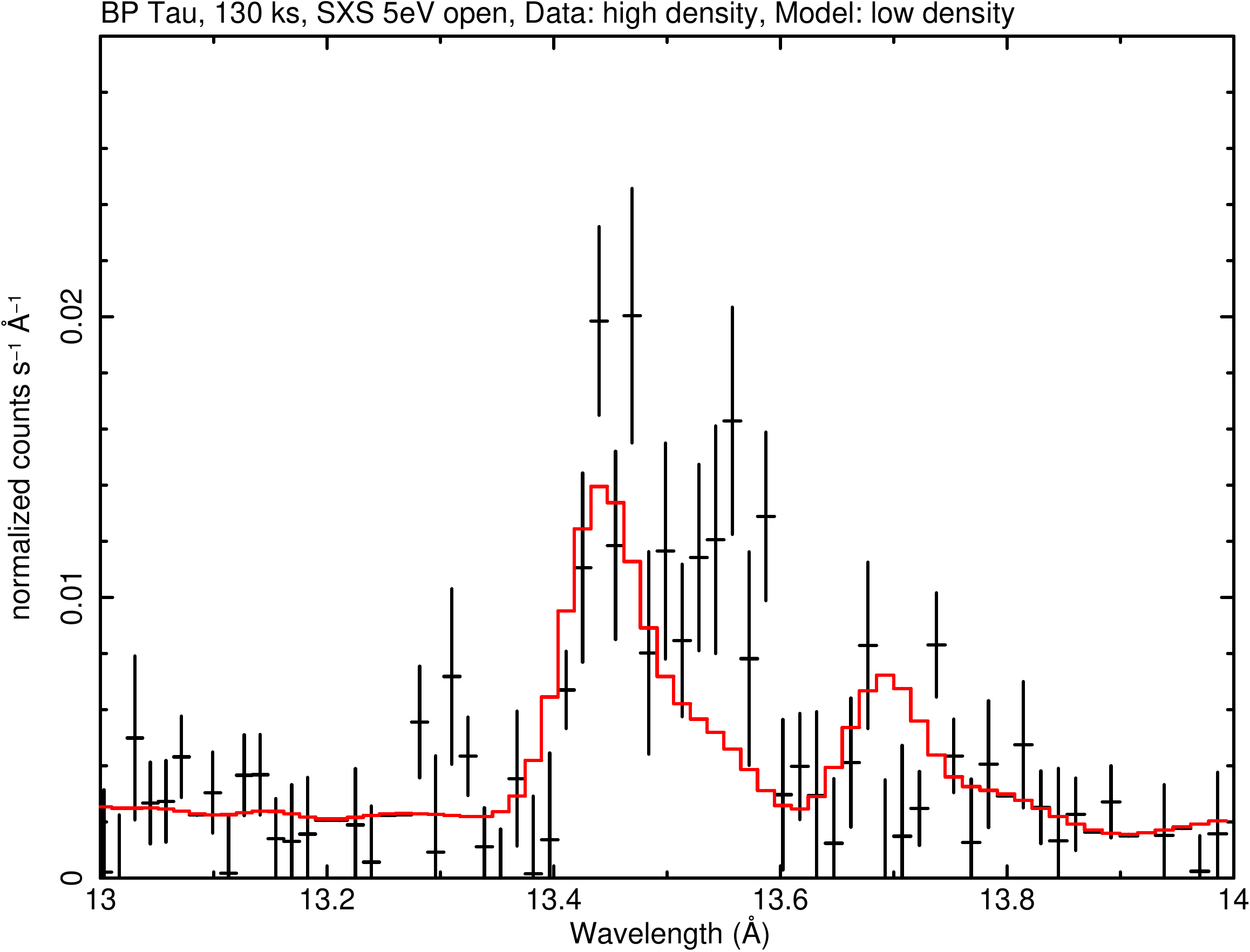}\hfill
\includegraphics[width=.44\hsize]{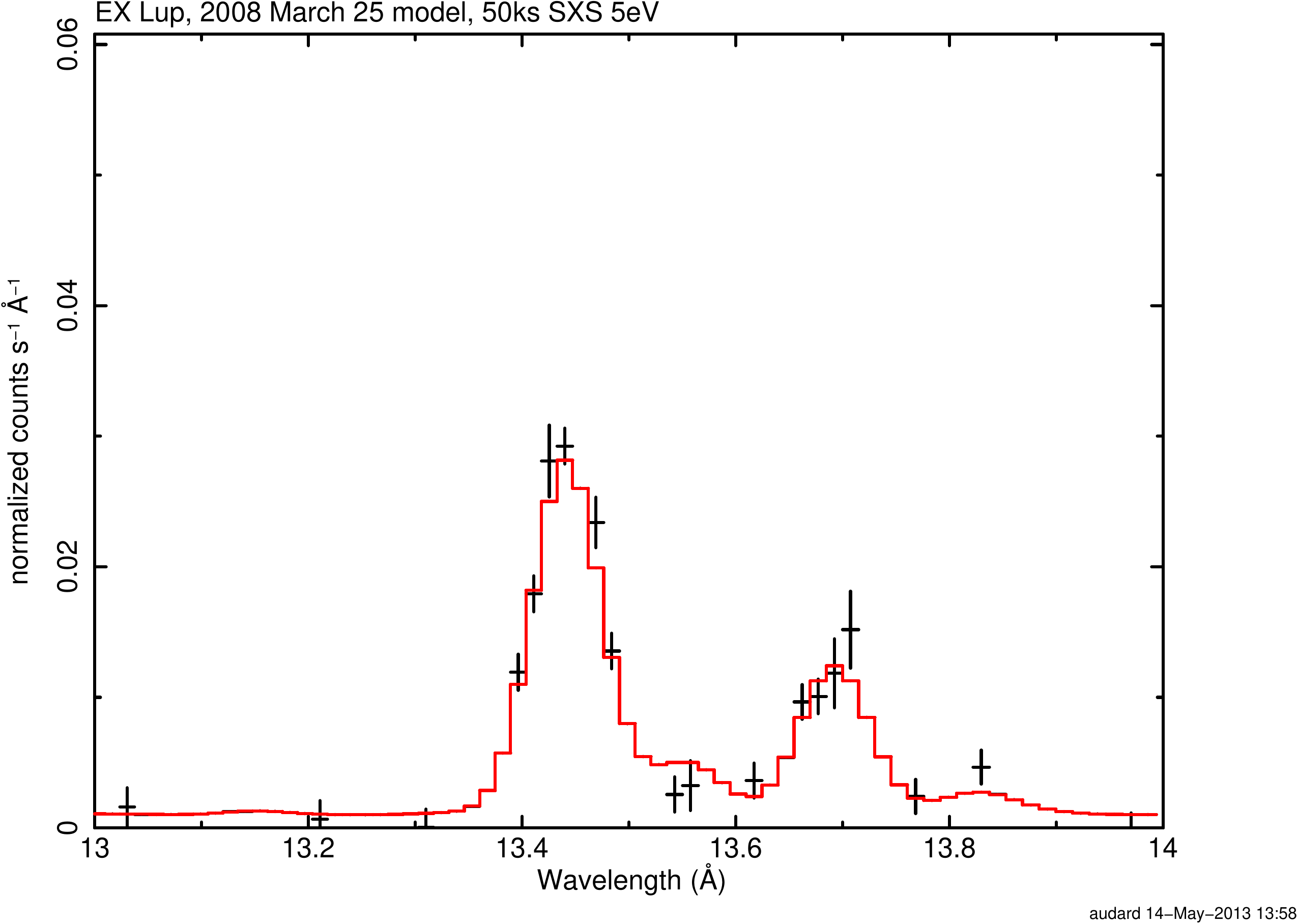}
\caption{{\it Left:} Simulation of a 130~ksec SXS observation of BP
  Tau. Both high and low density cases are presented as data points
  and a model curve, respectively \citep{schmitt05}.  {\it Right: }
  Simulation of a 50~ksec SXS observation of an outburst on EX Lup.  The model used
  in the simulation is based on the data obtained on 25 Mar 2008
  \citep{teets12}.}
\label{fig:ctts_sxs}
\end{figure}

Previous observations of young accreting stars with grating
spectrometers onboard {\it XMM-Newton} and {\it Chandra} were
rather rare (of the order of 15 targets), and many spectra had
relatively poor signal-to-noise ratios. As an example, a 130 ks
{\it XMM-Newton} observation of the nearby (140 pc) CTTS BP Tau
barely detected the O~{\sc vii} triplet, although it managed to hint
at an $f/i$ ratio indicative of high densities
\citep[Fig.~\ref{fig:bptau_xmm};][]{schmitt05}. The Ne~{\sc ix}
triplet, although better detected, also showed limited signal, whereas
the Mg~{\sc xi} or the Si~{\sc xiii} triplets were not detected.  In
contrast, {\it ASTRO-H} will be able to detect the Ne~{\sc ix}
(Fig.~\ref{fig:ctts_sxs}, left) and Mg~{\sc xi} triplets, and likely the
Si~{\sc xiii} triplet, with good signal-to-noise ratios and with much
better spectral resolution above 1~keV than with grating
spectroscopy\footnote{In terms of detecting weak lines, {\it ASTRO-H}/SXS
  has an advantage over {\it XMM-Newton}/RGS and {\it Chandra}/LETGS above
  0.6~keV.}. These lines will be crucial in probing the X-ray emission
of young stars for which accretion plays a major role: they
can probe densities of the order of $10^{11}$ to $10^{13}$~cm$^{-3}$,
a range of densities expected from X-ray emitting accretion
shocks. Furthermore, they will be less sensitive to photoelectric
absorption than the O~{\sc vii} triplet, which is often be severely absorbed in
young stars in star forming regions.

% Please note that A-H could be a poor satellite to be used for ToO. 
% For one, it lessens its efficiency; for another, because of its long
% platform length, it could take a very long time to repoint a telescope. 

In the case of outbursting sources, {\it ASTRO-H} will provide the {\it
  first} ever measurement of densities in such young accreting stars.
The observations, taken as soon as possible after the onset of the
outburst will allow measurement of changes in electron density.  In
the current view, accreting material would partly fill in the stellar
coronal loops and possibly change the magnetospheric configuration.
{\it ASTRO-H} will allow to directly determine the amount of material
accreted.  Little or no change in electron densities would, on the
other hand change dramatically the interpretation of the X-ray
emission and suggest that accretion events trigger magnetic
reconnection events in the young stellar corona.

{\it ASTRO-H} should ideally observe several ``normal''
accreting young stars to determine the electron densities from the
Ne~{\sc ix} and Mg~{\sc xi} triplets in a moderately-sized but
representative sample. A large sample would then provide a legacy survey of electron densities measured in
young accreting stars.

The second approach, dedicated toward outbursting sources should focus
on a ToO approach for EX Lup-type stars. The observation should be
triggered by ground-based observations in the optical and
near-infrared. The outburst should ideally be followed up by three to
five {\it ASTRO-H} observations, and should include post-outburst observation to
determine the conditions long after the accretion event.  Short pre-outburst observations could be
considered as well. In the case of FU Ori-type stars, an approach
similar to ``normal'' CTTS is sufficient in view of the very long time
scale of the outburst.

\subsection{Targets \& Feasibility}

The science topic is oriented toward two sets of accreting young stars.

First, accreting young stars in nearby star forming regions should be
observed. For example, the Taurus, Lupus, or Ophiuchus star forming
regions include several tens of potential targets that could be
observed. In view of the limited field-of-view of the SXS, the main
instrument to address the science goals of this topic, pointed
observations to select candidate members with promising X-ray spectra
based on CCD spectroscopy should be chosen. In particular sources with
soft X-ray components should be the prime targets.

Second, targets with known variable mass accretion rates displaying
episodic accretion events should be observed, at least twice, once
before an outburst, the second time in the early phase of the
outburst, after some trigger obtained from optical/near-infrared
ground-based observations. Although FU Ori-type targets would be
ideal, they generally are either too embedded or too faint,
notwithstanding the fact that a measurement in quiescence is unlikely
in the time frame of the {\it ASTRO-H} mission. Instead, EX Lup-type
objects, would be better suited. Several objects are possible, such as
EX Lup, V1118 Ori, V1647 Ori. Due to the impossibility to predict
which one of these targets would experience an accretion event after
the launch of {\it ASTRO-H}, a TOO approach, with an observation timescale
after trigger relatively flexible (several days to a few weeks are
sufficient, the outbursts lasting several months to years), would be
best. The second measurement could be derived either after the return
to quiescence, or a survey of 2-3 potential targets in quiescence could be done in the early phases of {\it ASTRO-H}.

Two feasibility examples of measurements of Ne IX triplets are shown in Figure~\ref{fig:ctts_sxs}.

\subsection{Beyond Feasibility}

Observation of a strong X-ray outburst of a young star due to an
accretion event is a definite bonus for this topic. In the present
strategy, a TOO approach is foreseen, and such an outburst is
therefore likely to be observed by {\it ASTRO-H}.
% if a trigger happens during the PV phase. 
The relatively frequent accretion activity observed in
V1647 Ori, V1118 Ori, EX Lup, and other EX-Lup-type outbursting
sources observed with {\it Chandra}, {\it XMM-Newton}, and
{\it Suzaku} shows that such an event is likely to occur.

A second bonus would be an accretion event in a ``normal'' accreting
CTTS, which would definitely show the link between the CTTS and the
outbursting sources.

A final bonus would be the detection of a strong flare: young stars
are also strongly magnetically active stars and should, therefore,
display flares. This is described in detail in Topic 2.

\clearpage

%%%%%%%%%%%%%%%%%%%%%%%%%%%%%%%%%%%%%%%%%%%%%%%%%%%%%%%%%%%%
\section{Colliding Winds in Massive Stars}
%%%%%%%%%%%%%%%%%%%%%%%%%%%%%%%%%%%%%%%%%%%%%%%%%%%%%%%%%%%

%\author{Prepared by Y. Sugawara and Y. Maeda}

\subsection{Background and Previous Studies}

Massive binaries composed of a Wolf-Rayet (WR) star, a Luminous Blue
Variable (LBV) star and an OB star often show higher temperature
plasma compared to other early-type stars due to colliding winds.
Ram-pressure balance between the two hypersonic winds determines the
position of the colliding-wind region. The shocked plasma is expected
to be heated to temperatures of 10$^{7}$--10$^{8}$~K.  Colliding-wind
binaries are ideal testing grounds for plasma-shock physics, mainly
because plasma properties vary with binary separation in many cases.

Colliding wind binaries are important laboratories for investigating
the physics of particle acceleration, since they provide access to
higher mass, radiation and magnetic field energy densities than those
in SNRs.  High-spatial-resolution radio observations of synchrotron
emission have revealed that the acceleration site is the region where
the winds collide \citep[e.g.,][]{dougherty05}.  In the X-ray band,
the {\it Suzaku} HXD detected excess emission in the hard (E $>$ 10 keV)
band from $\eta$ Car \citep[LBV$+$WR:][]{sekiguchi09} and WR 140
\citep[WR$+$O:][]{sugawara13}. However, because of limited counting
statistic in the HXD spectra, it is not possible to distinguish
whether the hard excess is thermal or non-thermal.

Mass-loss rate, stellar wind acceleration, and rotation velocity are
the most important and uncertain parameters in the evolution of a
massive star. The mass-loss rate and the acceleration are usually
measured using the radio and IR continuum flux together with modeling
of spectral lines in the optical and IR bands. However, independent
measurements of these parameters can be made in the X-ray band.  The
X-ray luminosity is highly dependent on the separation distance of the
binary, the mass-loss rates, and wind velocities (\citealt{stevens92};
\citealt{usov92}). If we know the orbital parameters of a binary, we
can infer the mass-loss rate and wind-acceleration from the
phase-dependent X-ray luminosity.

\begin{figure}[ht]
  \begin{minipage}{\textwidth}
    \begin{center}
      \includegraphics[width=\hsize]{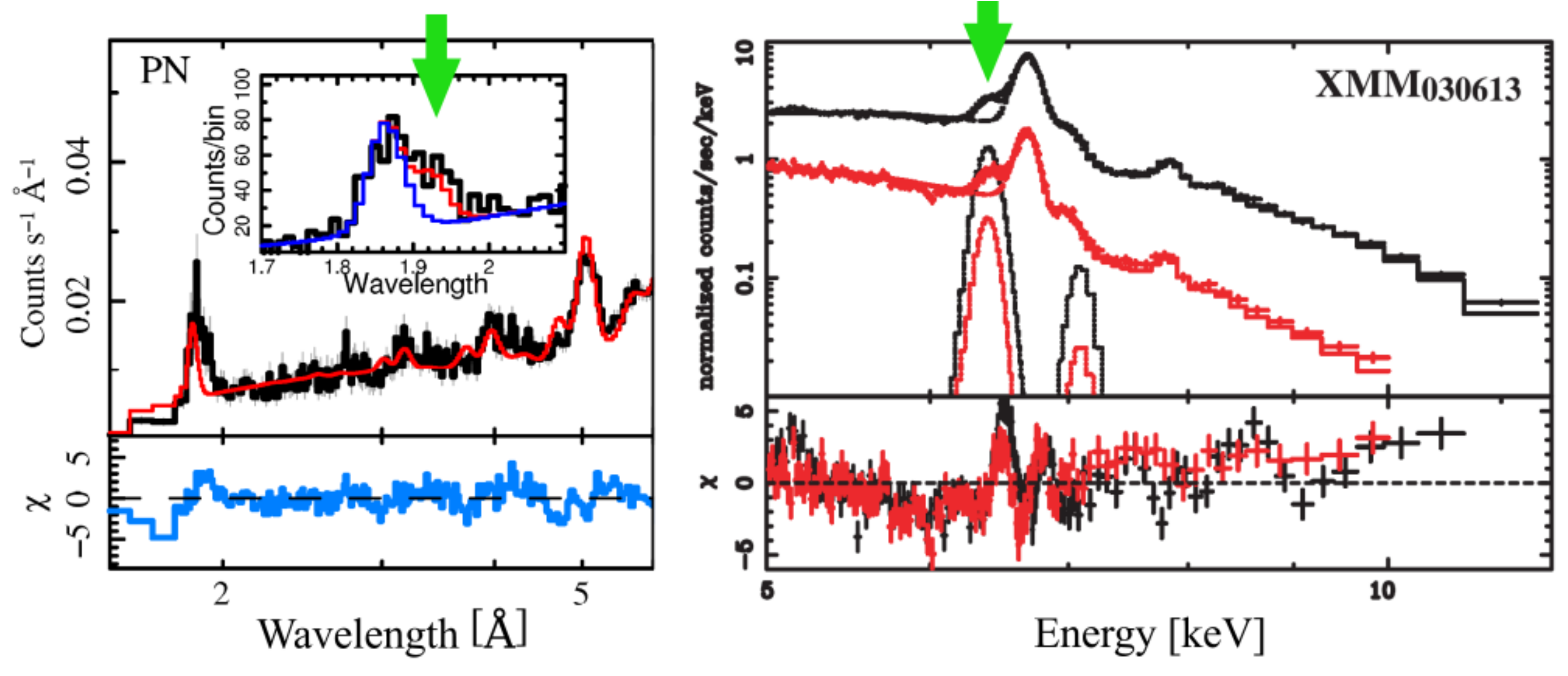}
    \end{center}
    \caption{Fluorescent neutral iron lines (green arrow). {\it Left
        panel}: Spectra of WR~6 above 2 keV (from Fig.2 in
      \citealt{oskinova12}). Black shows the data of {\it XMM-Newton}/EPIC-pn. An
      insert to the top in the panel shows a broad spectral feature that
      is not reproduced by the thermal plasma model (blue). An additional
      fluorescent Fe line at 1.92~\AA improves the fit (red). {\it Right
        panel}: Spectrum of $\eta$ Car above 5 keV (from Fig.8 in
      \citealt{hamaguchi07}). Black and red data points show {\it XMM-Newton}/EPIC-pn
      and {\it XMM-Newton}/EPIC-MOS1 spectra, respectively. The solid lines show the
      thermal plasma model together with additional fluorescent iron
      lines. \label{fig:CWBFeK}
    }
  \end{minipage}
\end{figure}

X-ray spectroscopy indicates that the profile of the neutral
fluorescent iron line at 6.4 keV may be used to reveal the geometry of
the wind-wind shock zone as well as wind and orbital parameters of
colliding wind binaries. The neutral fluorescent iron line has been
detected from two massive binaries (WR~6:\citealt{oskinova12}; $\eta$
Car:\citealt{hamaguchi07}; Fig~\ref{fig:CWBFeK}).  Unlike $\eta$~Car,
the spectrum of WR~6 shows a circumstellar absorption column density
too small to explain the observed equivalent width of the neutral
fluorescent line. The fluorescence must be emitted due to irradiation
of the surface of the secondary by hard X-rays from the post-shock
region, and the equivalent width therefore measures the solid angle
subtended by the secondary surface from the point of view of the
post-shock region.

\subsection{Prospects \& Strategy}

We describe two possible observational techniques with {\it ASTRO-H} in the
following section. First, we describe the use of broadband
spectroscopy to constrain the origin of the hard excess. Second, we
discuss the use of spectral lines to reveal the geometry of wind-wind
shock zones and measure wind and orbital parameters of CWBs.

\subsubsection{Broad-band spectroscopy}

The {\it ASTRO-H} HXI will characterize the origin of the hard X-ray excess
component, which may be related to $\gamma$-ray emission detected in
massive star clusters (e.g., Cyg OB2, IC1805, NGC 6383). We would like
to probe whether the hard excess is due to inverse Compton emission or
thermal bremsstrahlung.

First we consider interactions between relativistic electrons and
stellar UV photons \citep{bell78}. In the colliding wind region,
first-order diffusive shock acceleration results in the production of
a power-law spectrum of electrons with index 2 \citep{white95}.  The
luminosity ratio of inverse Compton to synchrotron radiation can be
written as ${L_{\rm{syn}}}/{L_{\rm{IC}}} = 840\, {\it
  B}^{2}{r^{2}}/{L_{\rm{bol}}}$, where the magnetic field ${\it B}$ is
expressed in \rm{G}, $r$ is the distance from the colliding-wind zone
to the star in {\rm AU}, and $L_{\rm bol}$ is the stellar luminosity
in solar units.

In the case of a strong adiabatic shock, the maximum electron
temperature is given by 
\begin{equation}
  k_{\rm{B}}T_{\rm e} = \frac{3}{16}\, \mu_{\rm w} m_{\rm p} v^{2}
  \sim 1.21 {\rm keV}\, \frac{\mu_{\rm w}}{0.62}\,
  \left(\frac{v}{1000~{\rm km~s}^{-1}}\right)^{2}\, ,
\end{equation}
where $\mu_{\rm w}$ is the mean atomic weight of the electron
plus ion mixture, $m_{\rm p}$ is the proton mass, and $v$ is the
pre-shock wind velocity (cf. \citealt{antokhin04}).  For example, if we
suppose $v=3000~{\rm km~s}^{-1}$, the post-shock temperature is $k_{\rm
B}T_{\rm e} \sim$ 18 $\mu_{\rm w}$~{\rm keV}.

Using SXS, SXI and HXI on {\it ASTRO-H}, we will for the first time be able
to differentiate between inverse Compton and thermal bremsstrahlung
radiation. 

\begin{figure}[ht]
  \begin{center}
    \begin{minipage}{0.9\textwidth}
       \includegraphics[width=\hsize]{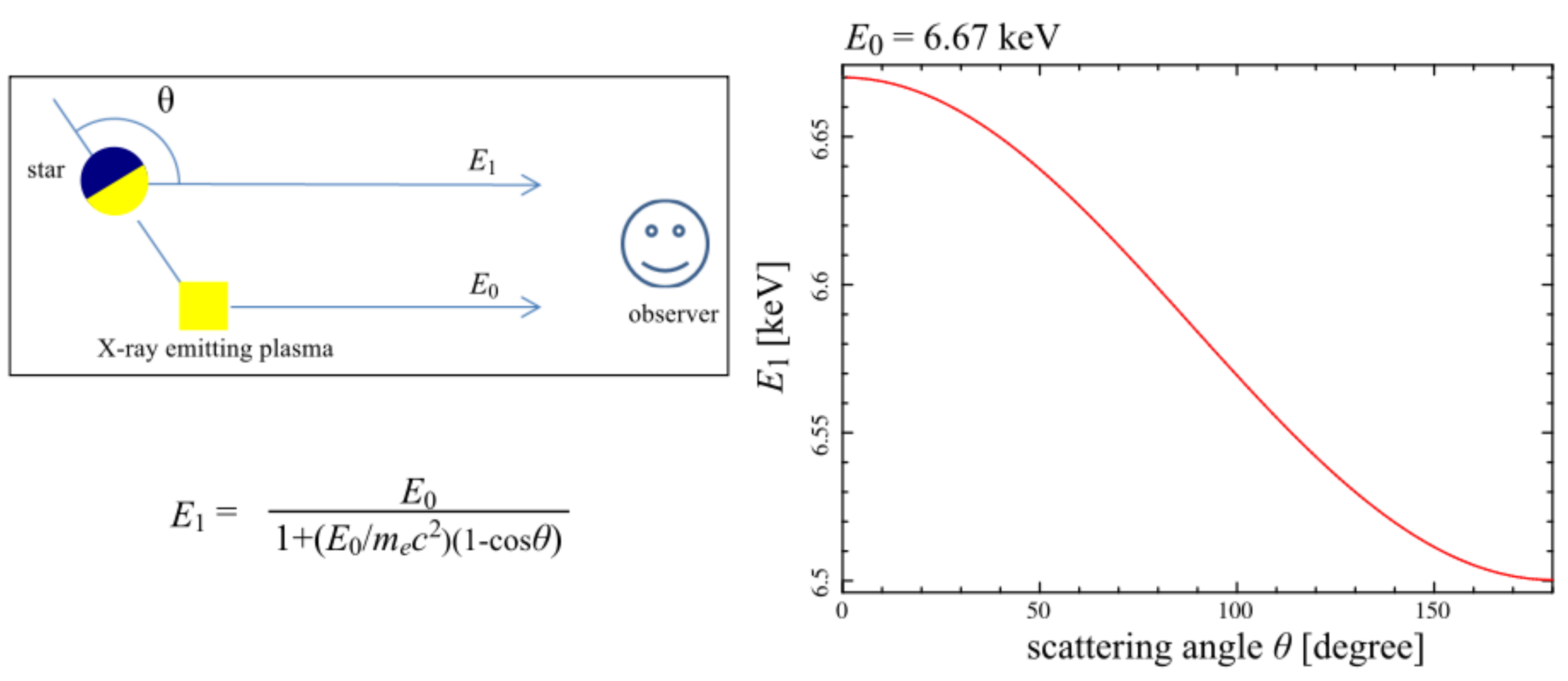}
      \caption{{\it Left panel}: A schematic view of Compton
        scattering. The equation shows the shift energy as a function of
        scattering angle. {\it Right panel}: The shift energy is shown as
        a function of scattering angle for rest energy $E_{0}$=6.67 keV.
        \label{fig:Compton}
      }
    \end{minipage}
  \end{center}
\end{figure}

\subsubsection{Spectral lines}

It is well known that the equivalent width of a neutral fluorescent
iron line is proportional to the solid angle subtended by the
reflector (e.g., \citealt{hayashi11}). For a colliding wind binary the
fluorescent iron emission line at 6.4 keV therefore probes the site of
the post-shock plasma, since the solid angle subtended by the
secondary surface is related to its separation from the post-shock
region; and by measuring the location of the post-shock plasma, we
also obtain the wind momentum ratio of the binary.

Detection of a shoulder due to Compton backscattering of lines would
further constrain the geometery (e.g., \citealt{watanabe03};
\citealt{mcnamara08}). The energy shift is related only to the
backscatter angle (Fig~\ref{fig:Compton}).

\subsection{Targets \& Feasibility}

\begin{figure}[ht]
  \begin{center}
    \includegraphics[width=0.7\hsize]{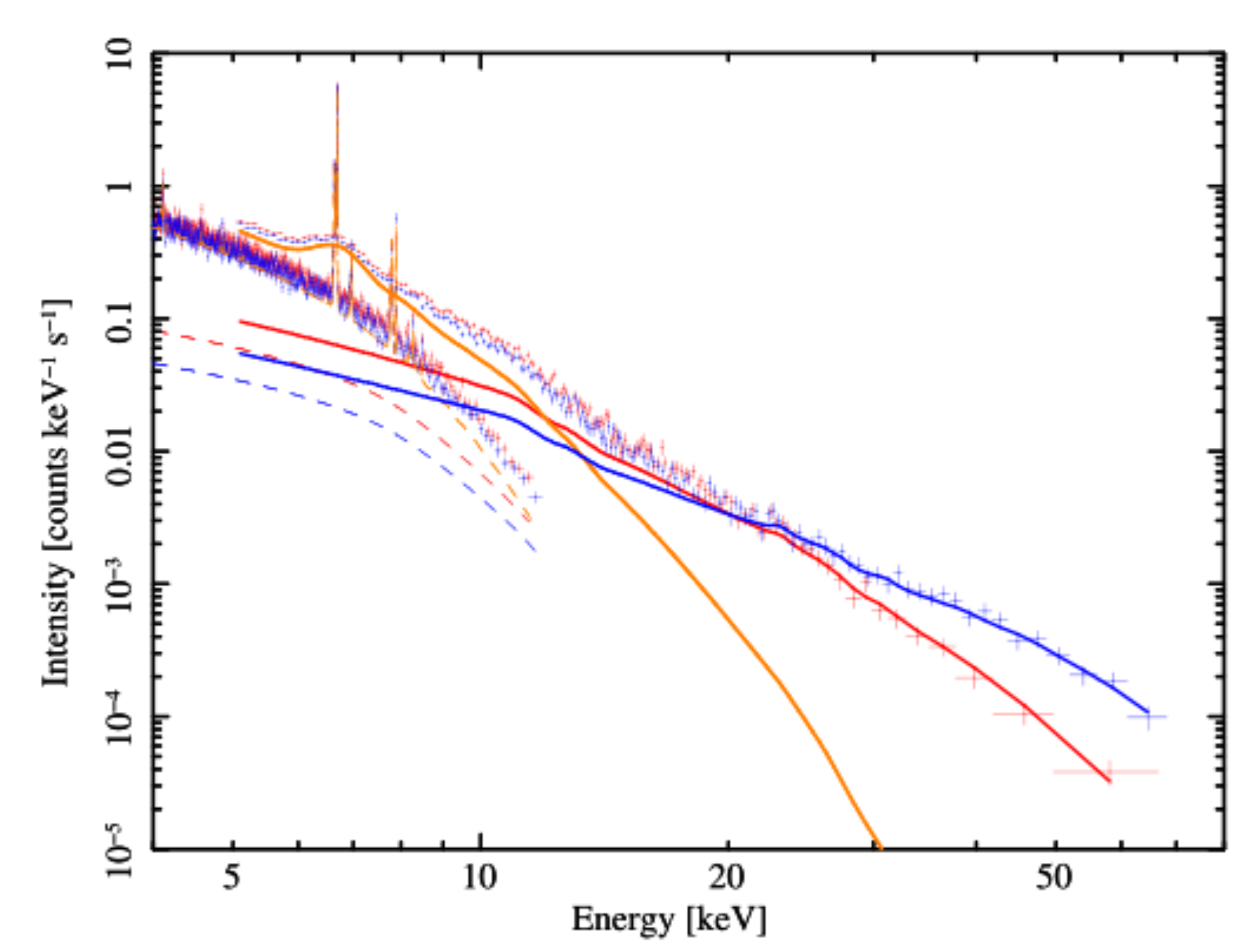}
  \end{center}
  \vspace{-0.8cm}
  \caption{Simulated SXS$+$HXI spectra of eta Car in a 50 ksec
    exposure. The blue and red curves show the inverse Compton and very
    hot thermal bremsstrahlung radiation components, respectively, with
    solid and dashed curves showing this model for HXI and SXS,
    respectively. The orange curve shows the thermal radiation from the
    colliding wind shock.}
  \label{fig:EtaCarSimulation}
\end{figure}

$\eta$ Car is one of the most suitable targets for {\it ASTRO-H}.  It is a
binary system \citep{damineli96} containing a very massive LBV (``the
primary'') and a hotter and less luminous evolved main-sequence
companion (``the secondary'').  $\eta$ Car is one of the brightest
colliding wind binaries in the X-ray band.  Previous {\it Suzaku} and
{\it XMM-Newton} obserations show evidence for a hard excess and
neutral Fe K fluorescent emission at 6.4 keV 
(\citealt{sekiguchi09}; \citealt{hamaguchi07}).

We show simulated SXS and HXI spectra of $\eta$ Car in a 50 ksec
exposure in Figure~\ref{fig:EtaCarSimulation}.  We simulated the
spectra for two cases: inverse Compton (blue); and thermal
bremsstrahlung radiation (red). These two cases can easily be
distinguished by the HXI.

\subsection{Beyond Feasibility}
Luminous Blue Variables are a rare sub-clas of massive star know to
undergo extreme eruption events. The primary of $\eta$~Car underwent
two major eruptions in the 19th century. The most extreme LBV
eruptions can result in mass loss of up to 10--40$M_{\odot}$ or more
(e.g., \citealt{smith06}). \cite{kashi10} speculated that major
eruptions are triggered by stellar companions and that in extreme
cases a short duration event with a huge mass transfer rate can lead
to a bright transient event on timescales of weeks to months (a
``supernova impostor''). Observations of eccentric massive binaries at
periastron passage could be an ideal opportunity to catch such an
eruption. Observation of such an event could help constrain supernova
theory.

In some CWBs, the wind momentum of the secondary may be insufficient
to prevent the wind of the primary from colliding directly with the
surface of the secondary, either for part or all of the binary orbit.
This may occur at periastron in $\eta$~Car, in which case it may be
possible to measure the rotation velocity of the secondary. It might
also be possible to make a similar measurement in systems where
neutral Fe K fluorescence from the surface of the secondary is
observed.  \cite{penny04} analyzed the UV line profiles of 177 massive
stars and suggested that the $v$ sin $i$ is at most 300 km s$^{-1}$,
which is within the capabilities of the SXS. Stellar rotation speeds
are critical ingredients in stellar evolution theory and in supernova
explosions, with more extreme rotators considered to be likely
candidates for Gamma-ray burst progenitors.

\clearpage

%%%%%%%%%%%%%%%%%%%%%%%%%%%%%%%%%%%%%%%%%%%%%%%%%%%%%%%%%%%%
\section{Diffuse X-rays Associated with Star Forming Regions}
%%%%%%%%%%%%%%%%%%%%%%%%%%%%%%%%%%%%%%%%%%%%%%%%%%%%%%%%%%%

\subsection{Background and Previous Studies}

Heavy elements such as carbon, oxygen, nitrogen, silicon and iron are
produced inside massive stars through nuclear fusion.  They are
released to interstellar space by stellar winds and supernova
explosions and enrich star forming clouds harboring young Suns.
Supernova explosions eject matter in one event, while stellar winds
from massive stars return more matter back to the interstellar medium
gradually throughout the life of the star \citep[e.g., Figure~6 of][]{Meynet2003}.
Both of these mechanisms contribute significantly to the chemical enrichment of the universe.

Transfer of heavy elements from massive stars to young stars can be witnessed in a massive star forming region (SFR) 
because massive stars evolve much more quickly than low mass stars.
The diffuse X-ray plasmas which have recently been discovered in many massive star
forming regions are suspected to originate from heated ejecta of massive stars, i.e., massive stellar winds and/or supernova remnants
(RCW38: \citealt{Wolk2002}, M17: \citealt{Townsley2003, Hyodo2008}, Carinae Nebula: \citealt{Hamaguchi2007a, Ezoe2008, Townsley2011},
Orion Nebula: \citealt{Gudel2008}).
Their physical parameters range from \KT\ $\sim$0.1--0.8~keV, log \LX\ up to 35~\UNITLUMI, and size of
$\sim$1--10$^{3}$~pc.  
They tend to fill cavities of dense, cold molecular clouds where newly born young stars are embedded.
Heavy elements in the plasmas may be absorbed into young solar systems through the cavity surfaces.

\begin{figure}
\begin{center}
\includegraphics[width=3in]{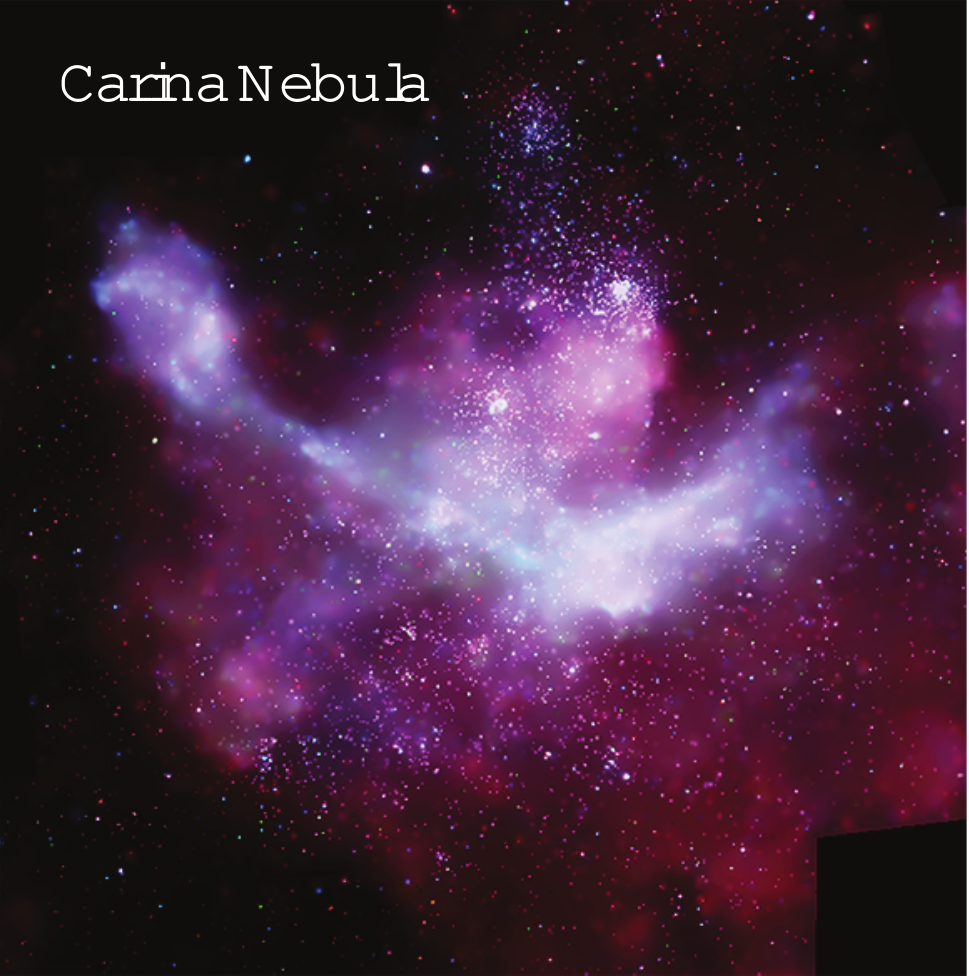}
\end{center}
\caption{X-ray emission from Carina \citep{Townsley2011}.}
\label{fig:carina_nebula_image}
\end{figure}

\begin{figure}
\begin{center}
\includegraphics[width=6in]{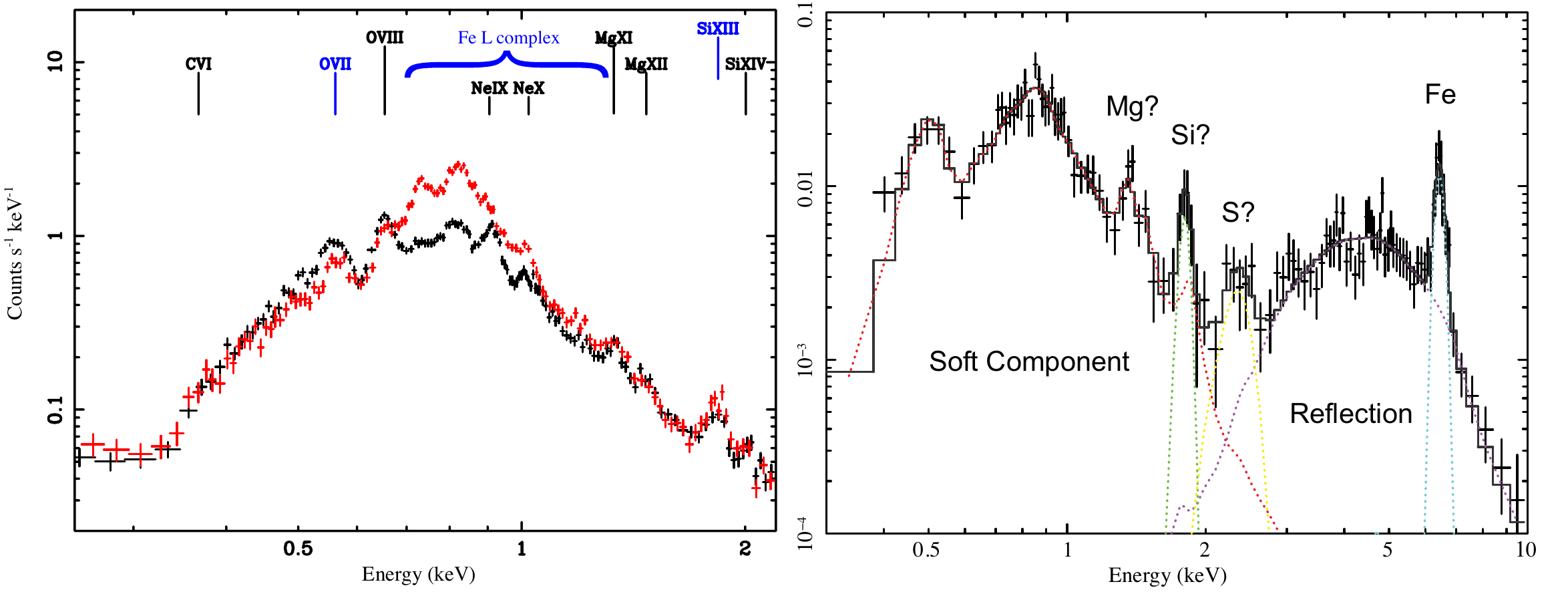}
\end{center}
\caption{
{\it Left:} Overlaid $0.3-2$ keV spectra from the north (black) and south (red) regions of \cite{Hamaguchi2007a} in Carina.
The north spectrum is normalized by 64\% to adjust for the relative area compared to the south region at 2~keV. 
The labels show detected emission lines; lines labeled in blue exhibit distinct differences between the north and south spectra.
{\it Right:} {\it Chandra} spectrum of the Homunculus nebula of \etacar (Hamaguchi et al., in prep.).
The spectrum shows peculiar strong emission lines around the silicon, sulfur and magnesium energies, which may
originate from a non-thermal process such as charge exchange in collision of hot stellar winds with ambient cold gas.
}
\label{fig:carina_nebula_spectra}
\end{figure}

It is still controversial whether this X-ray emitting plasma
originates primarily from supernovae or stellar winds.  Supernova
remnants (SNRs) such as Cas A, Kepler's remnant and the Cygnus loop
have hot plasmas with \KT $\approx$0.1--3~keV.  The strong spatial
gradient in elemental abundances observed from the Carina nebula
\citep[][see Figure~\ref{fig:carina_nebula_image},
  \ref{fig:carina_nebula_spectra} {\it left
    panel}]{Hamaguchi2007a,Townsley2011} is reminiscent of bullet-like
ejecta in SNRs which are rich in silicon, sulfur, etc.  However,
unlike conventional SNRs, none of the diffuse X-ray plasmas in star
forming regions has a spherical shape.  Furthermore, compact relics
expected from SNR explosions, black holes or neutron stars, have not
yet been observed in these regions, with the exception of a middle
aged neutron star found near the Carina Nebula
\citep{Hamaguchi2009,Pires2009}, whose origin is under discussion.
Even the Orion nebula shows diffuse X-ray emission, although it is
quite young, has only late O type stars, and is not thought to harbor
any SNR.

The Carina nebula also shows X-ray spectral features which are not
apparently explained by emission from collisional equilibrium (CE)
plasma.  \citet{Townsley2011} found excesses in the best-fit spectra
of many distinctive regions.  The excesses can be reproduced with
emission lines at $\sim$0.56, 0.76 and 1.85~keV, where no strong
emission lines from estimated plasma temperatures are present.  The
diffuse spectrum in the vicinity of \etacar\ also shows a peculiar
emission line at $\sim$1.79~keV between the energies of fluorescent
and Helium-like lines of silicon (Hamaguchi et al., in preparation,
right panel of Figure~\ref{fig:carina_nebula_spectra}).  Some SNRs
also show evidence for unidentified emission lines, which might be
related to the unidentified emission lines from the Carina nebula.
These lines may require strong Doppler shifts of nearby emission
lines, or non-thermal processes such as charge exchange.

The dynamics of and physical process in these diffuse plasmas are crucial in understanding their origin and interaction with star forming clouds.
We need to clearly resolve the emission lines around 0.5$-$1~keV with better than CCD resolution.  
An {\it XMM-Newton} grating  observation strongly constrained elemental abundances of the relatively concentrated ($\sim$1\ARCMIN) 
diffuse emission around \etacar \citep{Leutenegger2003}.
The \ah\ SXS will expand this pioneering study to more extended regions with peculiar X-ray features
and allow measurements of plasma dynamics and new physical processes
in extreme conditions.

\subsection{Prospects \& Strategy}

Diffuse X-ray plasmas in the SFRs, with typical temperatures of
$\lesssim$0.8~keV, emit K-shell emission lines of C, N, O, Ne, Mg and
Si, and L-shell lines of Si, S, Fe and Ni.  Among them, C, N, O, Si
and Fe are the key products of nuclear fusion and therefore good
probes to measure how stellar winds and/or supernova explosions inject
energy and metals into the interstellar medium.
These lines can be resolved clearly with the SXS and used for detailed diagnoses of
plasma temperature, elemental abundance and ionization state.

Optical emission and absorption line studies suggest the presence of high velocity gas of a few 100 to several 1000 \UNITVEL\
along the lines of sights to a few massive stars, as well as warm gasaround \etacar\ \citep{Walborn2002,Smith2008}.
This high velocity gas originates in flows from massive stars and/or from ancient supernova
explosions and is likely associated with thermally emitting X-ray plasma.
CCD resolution spectra have not detected any Doppler motions of the plasma above several 1000 \UNITVEL,
but high resolution SXS spectra will be able to trace velocity shifts and
broadening of these gases with much higher precision.

The SXS will clearly separate the heretofore unidentified emission
lines observed in CCD spectra from known thermal emission lines, and
hopefully detect more lines that are not expected from thermal plasma.
Their identifications will be securely made with measurements of their
line center energies at less than a few eV uncertainty.  These lines
will help to understand the physical processes and the origin of the
diffuse X-ray emission.\\

The SXS should start approaching this science with a few selected regions in one or
two SFRs with high X-ray surface brightness and which show interesting
spectral features in earlier X-ray observations and/or other
wavelengths.  A survey can reveal spatial variation in plasma dynamics
and elemental abundances in a SFR.  Better yet, multiple surveys of
SFRs may reveal the dependence of X-ray characteristics on the cloud
mass and age. A suitable candidate, for example, would be a pair of
regions in the Carina nebula, which show strong variations in the Fe
and Si elemental abundances \citep{Hamaguchi2007a,Ezoe2008,Townsley2011}.
The published {\it Chandra} survey of Carina 
\citep{Townsley2011} can serve as a guide 
in selecting regions with interesting unidentified X-ray emission line
features and which are dominated by diffuse emission with little
contamination from point sources.

\subsection{Targets \& Feasibility}

There are at least two distinct regions in the north and the south of
the Carina nebula.  Based on the {\it Chandra} studies of
\citet{Broos2011} and \citet{Townsley2011}, the best fields with high
surface brightness and low contamination from point sources are:
North: (R.A., Dec.) = (160.96070, -59.64274) (around the field
``inside 51'' in \citealt{Townsley2011}); South: (R.A., Dec.) =
(161.04080, -59.85487) (around the field ``inside 8'' in \citealt{Townsley2011}).
The combined point source flux between
0.5$-$2~keV in each of these fields is
$\sim$5$\times$10$^{-14}$\,\UNITFLUX, which is about $\sim$5\% of the
diffuse X-ray emission.

Other examples of SFR with diffuse X-ray emission are M17, RCW38,
Rosette Nebula, and M42.  They have significantly lower X-ray surface
brightness than the Carina nebula, so that the SXS observations would
require much longer exposure times ($\sim$ Msec).

\begin{figure}
\begin{center}
\includegraphics[width=\hsize]{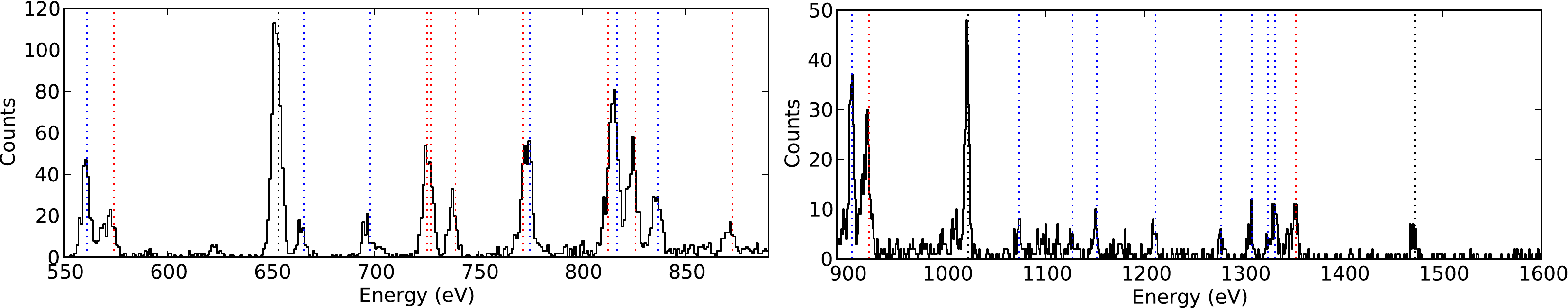}
\end{center}
\caption{ Simulated 30~ksec \ah/SXS X-ray spectrum of the central
  Carina nebula.  The model assumes that half of the emission
  originates from hot, ionized plasma, and the other half from charge
  exchange of the same population of ions with cold gas. ISM
  absorption of 2$\times$10$^{21}$~\UNITNH\ is included in the
  spectral model.  The blue and red dotted lines indicate the
  positions of emission lines which are expected to originate
  primarily from charge exchange and thermal plasma, respectively; the
  black dotted lines indicate emisison lines which are expected to be
  have strong emission from both mechanisms.  }
\label{fig:simulated_spectrum}
\end{figure}

%%%%%%%%%%%%%%%%%%%%%%%%%%%%%%%%%%%%%%%%%%%%%%%%%%%%%%%%%%%%
%\section{References}
%%%%%%%%%%%%%%%%%%%%%%%%%%%%%%%%%%%%%%%%%%%%%%%%%%%%%%%%%%%
%---------------------------------------------
% reference
%---------------------------------------------
\clearpage
\begin{multicols}{2}
\end{multicols}

\end{document}